%% file: manuscript_arXiv_v3.tex
\documentclass[prl, aps,twocolumn,superscriptaddress,longbibliography,floatfix,notitlepage]
{revtex4-1}

\usepackage{dcolumn}
\usepackage{bm}
\usepackage[utf8x]{inputenc}
\usepackage[english]{babel}
\usepackage{textgreek}
\usepackage[T1]{fontenc}
\usepackage{lmodern}
\usepackage{amsmath,amsfonts,amssymb,amsthm,bm,times,dcolumn}
\usepackage{mwe}
\usepackage{microtype}
\usepackage{braket}
\usepackage{mathrsfs}
\usepackage{nicematrix}
\usepackage{array}
\usepackage{tikz}
\usetikzlibrary{decorations.pathreplacing, arrows.meta}
\usepackage{blkarray}
\usepackage[colorlinks={true}, citecolor={blue}, filecolor={blue}, linkcolor={blue}, urlcolor={blue}]{hyperref}
\usepackage{graphicx,epstopdf,color}
\usepackage[mathscr]{euscript}
\usepackage{setspace}
\usepackage{physics}
\usepackage{pgffor} 

\begin{document}

\title{Control of Multipartite Entanglement through Anisotropy against Thermal Noise}

\author{Samudra Sur}
\email{samudra.sur@unige.ch}
\affiliation{Department of Quantum Matter Physics, University of Geneva, Quai Ernest-Ansermet 24, 1211 Geneva, Switzerland}

\author{Saikat Sur}
\thanks{author for correspondence}
\email{saikats@imsc.res.in}
\affiliation{Optics \& Quantum Information Group, The Institute of Mathematical Sciences, HBNI, CIT Campus, Taramani, Chennai 600113, India}


\begin{abstract}

Preserving multipartite entanglement in open many-body quantum systems is fundamentally limited by unavoidable environmental noise. We study the open-system dynamics of multipartite entanglement in an anisotropic XXZ spin chain interacting with a thermal spin bath, focusing on two states with   distinct types of multipartite entanglement: the  generalized Greenberger–Horne–Zeilinger and the generalized 
$W$ state. Using a master-equation approach combined with the Bethe ansatz technique, we show analytically that robustness of multipartite entanglement at low temperatures can be enhanced by suitably tuning the anisotropy of the system. Our results highlight interaction-induced spectral control as a mechanism for stabilizing multipartite entanglement in quantum computing platforms.

\end{abstract}

\maketitle

In  many-body quantum systems, \emph{ multipartite entanglement} serves as a key resource for tasks like quantum secret sharing~\cite{hillery_pra_1999,cleve_prl_1999,guhne_pr_2009,horodecki_rmp_2009},  quantum key distribution~\cite{epping_njp_2017,brunner_rmp_2014}, multiparty quantum teleportation~\cite{karlsson_pra_1998,chen_pra_2006}, quantum metrology~\cite{toth_pra_2012}, measurement-based quantum computation~\cite{raussendorf_prl_2001,briegel_natphys_2009}, quantum error-correction~\cite{gottesman_pra_1996,kitaev1997quantum}. Among the paradigmatic genuinely multipartite entangled states, the Greenberger–Horne–Zeilinger (GHZ) and the $W$ state represent two inequivalent classes of resource states that are distinct under local operations and classical communication (LOCC)~\cite{dur_2000}. A long-standing challenge in the development of quantum devices is preserving a quantum state with entanglement. However, initially pure states are prone to decoherence, and turn into mixed states, as complete isolation from the environment is practically unattainable~\cite{zeh1970,divicenzo_2000,zurek_rmp_2003}.
 
{\color{black} A generic mixed state is said to possess genuine multipartite entanglement (GME)  if it cannot be expressed as a mixture of states that are separable across any bipartition~\cite{shimoy_anyas_1995,acin_prl_2001}. However, in large many-body systems, quantification of GME is challenging as the number of bipartitions grows exponentially with system size~\cite{guhne_pr_2009,seevinck_pra_2008}. GME can become non-computable in the 
continuum limit~\cite{eisert_rmp_2010}, 
and unlike bipartite entanglement, GME 
has no unique quantitative measure for 
mixed states~\cite{horodecki_rmp_2009}. 
To address these difficulties, we use the 
\emph{distillation fraction} of a chosen 
reference family of resource states as a 
quantitative marker of GME, giving the 
asymptotic rate at which many copies of 
a state can be converted into the reference 
resource state under allowed local 
operations~\cite{dur_prl_1999,
horodecki_prl_2000}. For Dicke-type states, such as generalized $W$ states, GME measures can become non-extensive. We therefore employ the geometric measure of collective multipartite entanglement (CME) defined with respect to the set of fully separable states~\cite{wei_jmp_2010, shimoy_anyas_1995, barnum_jpha_2001}. Since the fully separable set is strictly contained within the biseparable set, this geometric measure constitutes a weaker entanglement criterion than other measures of GME~\cite{wei_jmp_2010}.}

{\color{black}While bipartite entanglement dynamics in 
open quantum systems have been fairly well 
studied~\cite{yu_prl_2004,clausen_pra_2012}, 
studies on collective entanglement in open 
quantum systems remain scarce,  either being
restricted to few-body 
systems~\cite{ma_njp_2013,vaishy_jpa_2022,
roy_mpla_2025} or non-interacting 
models~\cite{qiu_prb_2024,carvalho_prl_2004,
hashemi_pra_2012,roy_qreports_2022}. For 
interacting many-body systems, anisotropy 
can counterintuitively enhance robustness 
against thermal noise by regulating 
excitation gaps in the system spectrum, 
thereby suppressing thermal fluctuations 
from a broadband bath. The anisotropic XXZ 
spin chain provides a natural platform as the generalized GHZ and $W$ states appear 
in its eigenspectrum, and at $\Delta = 0$ 
the model maps to free fermions where 
entanglement is expected to degrade trivially. The 
anisotropy is thus the minimal interaction 
that breaks free-fermion symmetry and 
renders the decoherence dynamics 
non-trivial. Crucially, the model remains 
exactly solvable via the Bethe 
ansatz~\cite{bethe,izyu}, enabling fully 
analytical treatment in the thermodynamic 
limit, and is experimentally realizable 
in solid-state magnets and cold-atom 
platforms with externally tunable 
anisotropy~\cite{friedenauer_np_2008,
islam_nc_2011}.}

In this article, using the Master equation approach integrated with the Bethe ansatz, we analytically demonstrate that the open-system dynamics of multipartite entanglement, associated with these generalized GHZ and 
$W$ states is controlled by the anisotropy, particularly at low temperatures. The detailed calculations are provided in the Supplemental Information~\cite{supplement}, which  include the references~\cite{nielsen2000quantum, wei_pra_2003,nick, bastin_prl_2009,das_pra_2016,davis} not cited in the main text.  


  \begin{figure}[h]
\begin{center}
        \includegraphics[width=0.40\textwidth]{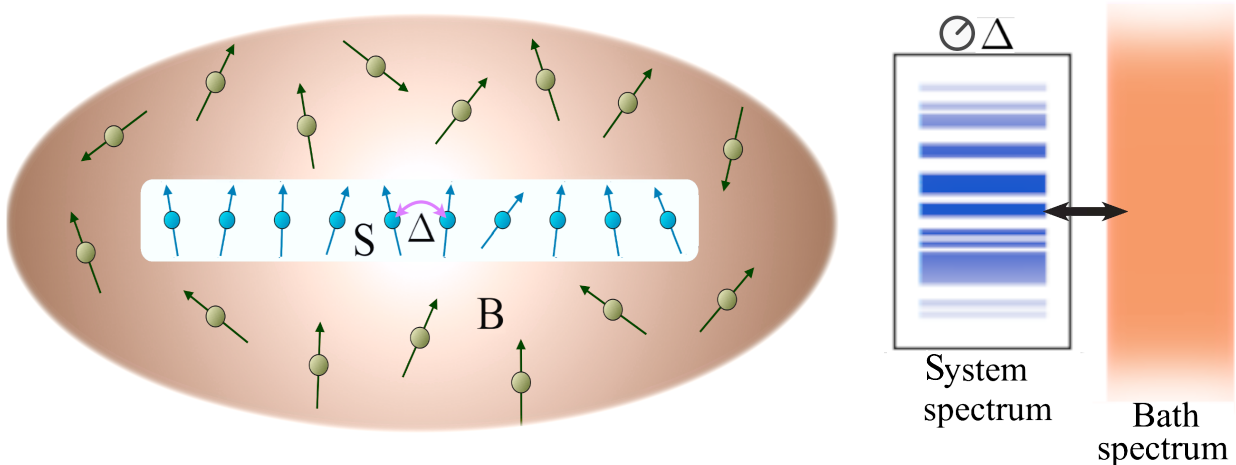}
\end{center}
\caption{{ Schematic of an interacting one-dimensional spin chain embedded in a spin bath. The intra-chain anisotropy can be externally tuned to control the multipartite entanglement against broadband thermal fluctuations.}}
\label{fig:schematic}
\end{figure}

\emph{Scenario}:  Using the Pauli matrices $\vec{\sigma}$ and $\vec{\varsigma}$ to represent the spin operators of the system and the bath, respectively, the Hamiltonian of the composite system is written as
$H = H_S + H_B + H_{SB}$, where 
\begin{align}
H_S &= -\frac{J}{2}\sum_{i=1}^{N}
\left( \sigma_i^x \sigma_{i+1}^x
+ \sigma_i^y \sigma_{i+1}^y
+ \Delta\, \sigma_i^z \sigma_{i+1}^z \right), \nonumber\\
H_B &= \sum_{k=1}^{N_b} \omega_k \varsigma^z_k,~~~
H_{SB} = \sum_{i=1}^{N}\sum_{k=1}^{N_b} g_{ik}
\left(\sigma^{+}_{i} \varsigma^{-}_{k} + \text{H.c.}\right),
\end{align}
denote the Hamiltonians of the system, the spin-bath, and the system--bath interaction, respectively. The anisotropy parameter $\Delta$ in $H_S$ denotes the interaction strength between spins aligned in the same direction. {\color{black}The exchange energy $J$ is set to unity throughout the manuscript.}
The states $\ket{0}$ ($\ket{1}$) are the eigenstates of the Pauli operator $\sigma^z$ with eigenvalues $+1$ ($-1$). A schematic of the system is shown in Fig.~\ref{fig:schematic}.

The system exhibits distinct quantum phases depending on the anisotropy: for $\Delta > 1$ and $\Delta < -1$, the ground state exhibits ferromagnetic and antiferromagnetic N\'eel order respectively with a finite excitation gap;  for $-1 < \Delta < 1$, the ground state is in a gapless critical phase.

The bath consists of $N_b$ non-interacting spin-$\tfrac{1}{2}$ particles, with $\omega_k$ denoting the transition frequency of the $k$~th bath spin, distributed isotropically and homogeneously in space~\cite{pra_2011}.
Each bath spin is  prepared in a thermal state
$\rho_k = p(\beta,\omega_k)|0\rangle\langle 0| + [1 - p(\beta,\omega_k)] |1\rangle\langle 1|$ ,
where the population
$p(\beta,\omega_k) = e^{-\beta \omega_k}/[2\cosh(\beta \omega_k)]$
corresponds to a Gibbs distribution at inverse temperature $\beta$ {\color{black}($\beta^{-1}$ is the temperature in units of $J/k_B$)}. {\color{black}We further assume that the bath has a 
broad, dense frequency spectrum acting 
as a thermal reservoir. The system-bath 
coupling $g_{ik}$ is weak and bilinear, 
allowing slow exchange of excitations 
between the system and the bath via a 
flip-flop interaction~\cite{petruccione}.  We work in the thermodynamic limit where 
both $N$ and $N_b$ tend to infinity with 
fixed finite ratio $N/N_b \equiv f < 1$.}

The Hilbert space of the system decomposes into magnetization sectors labeled by the number of down spins $ l$  relative to the fully polarized state $\vert 0 \rangle^{\otimes N}$. A basis for the $l$-magnon subspace is given by the ordered configurations 
$\ket{\mathbf{x}_l} \equiv \ket{x_1, x_2, \ldots, x_l}$,
where \( x_j \) denotes the lattice position of the \( j \)-th down spin. The corresponding eigenstates are $\vert \Psi_l(\mathbf{q}_l) \rangle = \sum_{\mathbf{x}_l} 
\psi(\mathbf{q}_l \vert \mathbf{x}_l) \, \vert \mathbf{x}_l \rangle$,
where \( \psi(\mathbf{q}_l \vert \mathbf{x}_l) \) is the Bethe wavefunction amplitude and the set $\mathbf{q}_l = (q_1, q_2, \ldots, q_l)$ are the Bethe roots or quasi-momenta that are obtained by solving the Bethe ansatz equations~\cite{bethe,izyu,sur_2017}. 

\par
The dynamics of the system in the presence of the bath is described, under weak system-bath coupling, by the second-order time-convolutionless (TCL) master equation.{\color{black} In the interaction picture, the reduced 
density matrix (RDM) of the system, 
$\rho_S(t)$, obeys~\cite{petruccione}
\begin{align}
\dot{\rho}_S(t) = -\int_0^t d\tau\, 
\mathrm{Tr}_B \!\left(
\left[ H_{SB}(t), \left[ H_{SB}(t-\tau), 
\rho_S(t)\otimes\rho_B \right] \right]
\right),
\label{eq:ME}
\end{align}
where $H_{SB}(t) = e^{i (H_S + H_B) t}\, 
H_{SB}\, e^{-i (H_S + H_B) t}$ denotes 
the system-bath coupling Hamiltonian in 
the interaction picture with respect to 
the total free Hamiltonian $H_S+H_B$, 
and $\rho_B$ is the equilibrium density 
matrix of the bath.}
 For a translationally invariant system coupled to 
an isotropic and homogeneous environment, the couplings satisfy 
$g_{ik} = g(k)$, ensuring that the flip-flop interaction $H_{SB}$ cannot 
mix momentum sectors, and total momentum is thereby conserved in all 
open-system interaction processes~\cite{saikat_graph, petruccione}.


\emph{Generalized GHZ state:} We start by looking at the evolution and the decoherence of the generalized GHZ state in the presence of the bath. This state $\left(\vert 0 \rangle^{\otimes N} + \vert 1 \rangle^{\otimes N}\right)/\sqrt{2}$ is an equal coherent superposition of the zero-magnon and $N$-magnon Bethe eigenstates. {\color{black}Since the bath coupling changes the magnon 
number by $\pm 1$, the 
dynamics generated from this initial state is 
confined to the restricted Hilbert space 
spanned by the zero-, one-, $(N-1)$-, and 
$N$-magnon sectors. Consequently, the 
effective dynamics is captured by a 
$4\times 4$ density matrix in the  basis 
$\{ \vert \Psi_0\rangle, \vert \Psi_{1} 
(q=0)\rangle, \vert \Psi_{N-1} (q=0) \rangle,  
\vert \Psi_N\rangle \}$. The time-evolved generalized GHZ state in the Bethe basis reads as (for details, see Section I in \cite{supplement})}
\begin{subequations}
\begin{eqnarray}
  \rho{(t)} = \begin{pmatrix}
      \frac{u}{2} & 0 & 0 & v\\
      0 & \frac{1-u}{2} & 0 & 0\\
      0 & 0 & \frac{1-u}{2} & 0\\
      v & 0 & 0 & \frac{u}{2}
\label{eq: GHZ_evolution}   \end{pmatrix}, 
\end{eqnarray}where the time-dependent population (diagonal) and coherence (off-diagonal) terms are obtained by solving the ME in Eq.~(\ref{eq:ME})
{\color{black}
\begin{eqnarray}
  &&u(t) =
  1 -
  \frac{e^{-\beta(\Delta - 1)}}{2\cosh[\beta(\Delta - 1)]}
  \left(1-e^{-\pi t\gamma f n}\right),
  \nonumber\\
  &&v(t) =
  \frac{1}{2}
  \exp\left[
  -\pi t\gamma f n\,
  \frac{e^{-\beta(\Delta - 1)}}{2\cosh[\beta(\Delta - 1)]}
  \right].
  \label{eq: u_v}
\end{eqnarray}
}
\end{subequations}
Here \(n\) denotes the density of states for the bath spins, and $\gamma$ is an effective decay parameter that depends on several microscopic features of the composite system, including the spatial distribution of bath spins and the dimensionality of the surrounding environment (refer to Section II in \cite{supplement} for the explicit form of $\gamma$). The temperature and the anisotropy together dictate the dissipation rates through the Gibbs factor. Population transfer from the state \(\ket{\Psi_0}\) to \(\ket{\Psi_1(q=0)}\) (or from \(\ket{\Psi_N}\) to \(\ket{\Psi_{N-1}(q=0)}\)) occurs on a time scale inversely proportional to the product \(\gamma f n\), while the coherence decays on a time scale inversely proportional to \(\gamma f n\) multiplied by a temperature-dependent Gibbs factor.

Both the population and coherence terms in Eq.~\eqref{eq: u_v} exhibit a non-analytic behavior at \(\Delta = 1\) in the low-temperature limit. This non-analyticity originates from the exponential factor {\color{black}\(\exp\{-\beta(\Delta-1)\}\)}, as its asymptotic behavior changes discontinuously as \(\Delta \to 1^\pm\). This exponential factor originates from the transition rate connecting the zero-magnon and one-magnon sectors. As a result, the dissipative dynamics differs qualitatively between the regimes \(\Delta < 1\) and \(\Delta > 1\). {\color{black}In the low-temperature regime, both population exchange and coherence decay are strongly enhanced for \(\Delta < 1\), whereas the state remains comparatively robust for \(\Delta > 1\)}.

{\color{black} The 
bath can only inject or remove a single magnon at a time. The rate of these transitions is controlled 
by the Boltzmann weight of the energy gap 
$(\Delta-1)$ between the zero and 
one-magnon sectors. For $\Delta > 1$, this 
energy gap is positive and thermally suppressed 
at low temperature, protecting the GHZ 
coherence. However, for $\Delta < 1$, the gap is negative, causing transitions thermally 
favorable and rapid decoherence 
at any nonzero temperature. The anisotropy of the system thus tunes 
the Boltzmann suppression of magnon 
excitations, and  directly controls the 
survival of GHZ-type multipartite 
entanglement .This non-analyticity in the dissipative rates leads to sharp features in time-dependent entanglement measures across \(\Delta = 1\), as we shall discuss below.}


We first look at the \textit{geometric measure} to characterize the GME of the state in Eq.~\ref{eq: GHZ_evolution}. For a generic mixed state of the form $\rho = \sum_i p_i \ket{\psi_i}\bra{\psi_i}$ it is defined as the convex-roof minimization of the GME of its pure state components as
\begin{equation}
E_G(\rho)
\equiv \min_{\{p_i,\ket{\psi_i}\}}
\sum_i p_i E_G(\ket{\psi_i}),
\label{eq:convexity}
\end{equation}
{\color{black}where the minimum is over all possible 
choices of probabilities $\{p_i\}$ and 
pure states $\{\ket{\psi_i}\}$ decomposition of
$\rho$, and the pure state GME measure 
$E_G(\ket{\psi}) \equiv 1 - 
\max_{|\Phi\rangle}|\langle\Phi|\psi\rangle|^2$ 
is computed by maximization over all pure 
biseparable states 
$\ket{\Phi}$~\cite{shimoy_anyas_1995,
barnum_jpha_2001,das_pra_2016,wei_pra_2003}. In contrast, bipartite entanglement measures are defined for a specific bipartition}.  For
a state with incoherent mixture of orthogonal components supported on different excitation-number sectors of the form Eq.~\eqref{eq: GHZ_evolution}, a closed form  expression for the upper-bound can be evaluated directly from computing the GME of its  components as $E_G(\rho) \le \tfrac{u}{2}E_G(\rho_{GHZ}) + (1-u) E_G(\ket{\Psi_1(q=0)})$. 
The first term on the right hand side of the inequality appears from the zero-magnon and $N$-magnon sector; whereas the second term from the one-magnon and $(N-1)$-magnon sector. Now, it can be shown that the geometric GME of the generalized $W$ state is non-extensive, i.e., it scales as $N^{-1}$ with the system size $N$ (see Section IVB in \cite{supplement} for details). Therefore, in the thermodynamic limit, the contribution from the single-magnon and $(N-1)$-magnon sectors vanishes, and the expression becomes exact~\cite{wei_pra_2003,smolin_pra_2005,hubener_pra_2009,eltschka_jpha_2014}(see Section IVA in \cite{supplement}) in terms  of $u$ and $v$ (Eq.~\eqref{eq: u_v})
\begin{eqnarray}
  E_G(\rho) = \frac{u}{2} \left( 1- \sqrt{1- \frac{4v^2}{u^2}} \right).
  \label{eq: GHZ_geometric_2}
\end{eqnarray}

{\color{black} Since $v(t)$ decays exponentially at a rate 
$\pi\gamma f n \cdot p(\beta,\Delta-1)$, and $u(t)$ relaxes  toward its thermal value, at early times or low 
temperatures with $\Delta > 1$, $v \approx 
u/2$ and $E_G \approx u/2$, recovering 
near-maximal GHZ entanglement. As $v \to 0$ at long times 
or high temperatures, $4v^2/u^2 \to 0$ and 
$E_G \to 0$, the state loses all GHZ-type 
GME. For $\Delta > 1$ at low temperature, 
$p(\beta,\Delta-1)$ is exponentially 
suppressed, slowing the decay of $v$ and 
protecting the GME. The ratio $v/u$ thus 
serves as the natural order parameter for the GME.}


Although the geometric measure of GME provides a  definition  of nonclassical correlations without reference to any specific task \cite{wei_pra_2003, huber_pra_2013}, the distillation fraction is an operational and protocol-dependent notion, quantifying how much useful GME can be extracted via  LOCC \cite{bennett_pra_1996, bennett_prl_1996,murao_pra_1998, dur_prl_1999,murao_pra_1999,clausen_2003,dur_ropp_2007,horodecki_rmp_2009}. Consequently, a nonzero geometric measure does not guarantee a finite distillation, particularly in the thermodynamic limit. We adopt Murao et al.’s protocol since it directly distills GHZ-type multipartite entanglement under LOCC, providing a simple computable distillation rate, and is straightforward to generalize for large many-body systems~\cite{murao_pra_1998}. 


{\color{black} The GHZ distillation protocol extracts 
perfect $N$-qubit GHZ states from many 
imperfect copies of $\rho$ via multipartite 
LOCC. Two identical copies serve as control 
($C$) and target ($T$) registers. Each 
party $j$ applies a local CNOT from $C_j$ 
to $T_j$, transferring bit-flip errors to 
the target. A collective even-parity 
measurement on $T$ retains only outcomes 
in $\ket{0}^{\otimes N}$ or 
$\ket{1}^{\otimes N}$, after which $T$ is 
traced out. Iterating $r$ times suppresses 
Dicke-type one-magnon contributions by 
$1/\sqrt{N}$ per round, leaving the state 
in the GHZ-diagonal subspace. A final 
hashing step addresses residual phase 
errors (see Fig.~\ref{fig:distillation_circuit}). 
The asymptotic lower bound on the 
distillable GHZ entanglement in the 
thermodynamic limit is (see Section~III 
in \cite{supplement})},
\begin{eqnarray}
 \mathcal{E}^{\text{GHZ}}_D = \max_{r \geq 0} \left[u^{(2^{r+1} - 2)}  \left(1  - h\left(P^r_0,P^r_1\right) \right) \right],  
\end{eqnarray}
with $h\left(P^r_0,P^r_1\right)$ denoting the Shannon entropy of the marginal phase-error distribution of the state at the $r~$th iteration. {\color{black} The first factor $u^{(2^{r+1}-2)}$ 
is the cumulative success probability of 
the $r$ iterative CNOT and postselection 
rounds, and the second factor 
$\left(1 - h\left(P^r_0,P^r_1\right)\right)$ 
is the hashing rate.  Maximizing over $r$ balances the exponentially decreasing success 
probability against the exponentially 
improving GHZ purity.}


\begin{figure}[h]
\centering
\begin{tikzpicture}[
    scale=0.54,
]

\def\xin{0}
\def\xcnotA{2.2}   
\def\xcnotB{3.0}   
\def\xcnotC{3.8}   
\def\xparity{5.8}  
\def\xstab{8.2}    
\def\xout{10.2}    

\def\Ca{1.5}    
\def\Cb{0.9}    
\def\Cc{-0.8}   

\def\Ta{-2.5}   
\def\Tb{-3.1}   
\def\Tc{-4.8}   

\node[left] at (\xin,\Ca) {$C_1$};
\node[left] at (\xin,\Cb) {$C_2$};
\node[left] at (\xin,\Cc) {$C_N$};
\node[left] at (\xin,\Ta) {$T_1$};
\node[left] at (\xin,\Tb) {$T_2$};
\node[left] at (\xin,\Tc) {$T_N$};

\node at (\xin+0.3, 0.5*\Cb+0.5*\Cc) {$\vdots$};  
\node at (\xin+0.3, 0.5*\Tb+0.5*\Tc) {$\vdots$};  
\node at (\xout+0.2, 0.5*\Cb+0.5*\Cc) {$\vdots$}; 

\draw (\xin,\Ca) -- (\xstab-0.55,\Ca);
\draw (\xin,\Cb) -- (\xstab-0.55,\Cb);
\draw (\xin,\Cc) -- (\xstab-0.55,\Cc);

\draw (\xin,\Ta) -- (\xparity-0.5,\Ta);
\draw (\xin,\Tb) -- (\xparity-0.5,\Tb);
\draw (\xin,\Tc) -- (\xparity-0.5,\Tc);

\draw[dashed, gray]
    (\xin-1.8, -1.65) -- (\xparity+0.6, -1.65);

\draw[decorate, decoration={brace, amplitude=5pt}]
    (\xin-1.7, \Cc-0.25) -- (\xin-1.7, \Ca+0.25)
    node[midway, left=6pt] {\small $C$};
\draw[decorate, decoration={brace, amplitude=5pt}]
    (\xin-1.7, \Tc-0.25) -- (\xin-1.7, \Ta+0.25)
    node[midway, left=6pt] {\small $T$};


\filldraw (\xcnotA,\Ca) circle (3pt);
\draw[thin] (\xcnotA,\Ca) -- (\xcnotA,\Ta);
\draw (\xcnotA,\Ta) circle (7pt);
\draw (\xcnotA-0.18,\Ta)--(\xcnotA+0.18,\Ta);
\draw (\xcnotA,\Ta-0.18)--(\xcnotA,\Ta+0.18);

\filldraw (\xcnotB,\Cb) circle (3pt);
\draw[thin] (\xcnotB,\Cb) -- (\xcnotB,\Tb);
\draw (\xcnotB,\Tb) circle (7pt);
\draw (\xcnotB-0.18,\Tb)--(\xcnotB+0.18,\Tb);
\draw (\xcnotB,\Tb-0.18)--(\xcnotB,\Tb+0.18);

\filldraw (\xcnotC,\Cc) circle (3pt);
\draw[thin] (\xcnotC,\Cc) -- (\xcnotC,\Tc);
\draw (\xcnotC,\Tc) circle (7pt);
\draw (\xcnotC-0.18,\Tc)--(\xcnotC+0.18,\Tc);
\draw (\xcnotC,\Tc-0.18)--(\xcnotC,\Tc+0.18);

\node at (0.5*\xcnotB+0.5*\xcnotC, 0.5*\Cb+0.5*\Cc) {$\vdots$};
\node at (0.5*\xcnotB+0.5*\xcnotC, 0.5*\Tb+0.5*\Tc) {$\vdots$};

\draw[thick, fill=yellow!20]
    (\xparity-0.75, \Tc-0.40)
    rectangle
    (\xparity+0.75, \Ta+0.40);
\node[align=center] at (\xparity, 0.5*\Ta+0.5*\Tc)
    {\small Parity\\[-2pt]\small meas.};

\draw[double, ->]
    (\xparity, \Tc-0.35)
    -- (\xparity, \Tc-1.0)
    node[below, align=center]
    {\small post-select\\[-2pt]
     \small $\ket{0}^{\otimes N},\ket{1}^{\otimes N}$};

\node[right] at (\xparity+0.8, 0.5*\Ta+0.5*\Tc)
    {\small (traced out)};

\draw[thick, fill=green!15]
    (\xstab-1.10, \Cc-0.35)
    rectangle
    (\xstab+1.10, \Ca+0.35);
\node[align=center] at (\xstab, 0.5*\Ca+0.5*\Cc)
    {\small GHZ\\[-2pt]\small stabilizer\\[-2pt]\small meas.};

\draw[double, ->]
    (\xstab, \Cc-0.35)
    -- (\xstab, \Tc-0.01)
    node[below, align=center]
    {\small $h\left(P^l_0,P^l_1\right)$};

\draw (\xstab+1.10,\Ca) -- (\xout,\Ca);
\draw (\xstab+1.10,\Cb) -- (\xout,\Cb);
\draw (\xstab+1.10,\Cc) -- (\xout,\Cc);

\node[above] at (0.8, \Ca+0.3)
    {\small $\rho^{(l)}\otimes\rho^{(l)}$};
\node[above] at (\xout-0.3, \Ca+0.3)
    {\small $\rho^{(l+1)}$};

\draw[->, thick, dashed]
    (\xout, \Ca+0.15)
    .. controls (\xout+0.5, \Ca+1.8)
             and (-0.8,     \Ca+1.8)
    .. (\xin, \Ca+0.15)
    node[midway, above] {\small repeat $r$ times};

\end{tikzpicture}
\caption{Quantum circuit for round $l$ of the $N$-qubit GHZ 
distillation protocol for $N$ qubits. Both the 
control register ($C$)  the target register ($T$) are initialized in the identical state $\rho^{(l)}$. The entire target register then undergoes a collective 
even-parity measurement projecting onto 
$\ket{0}^{\otimes N}$ or $\ket{1}^{\otimes N}$. Only successful outcomes are 
retained, and the post-selected control register becomes the 
input $\rho^{(l+1)}$ to the next iteration. After $r$ iterations 
the state converges to the GHZ-diagonal form given in Section IIIB in \cite{supplement}.}
\label{fig:distillation_circuit}
\end{figure}

\begin{figure}
\includegraphics[width=0.48\textwidth]{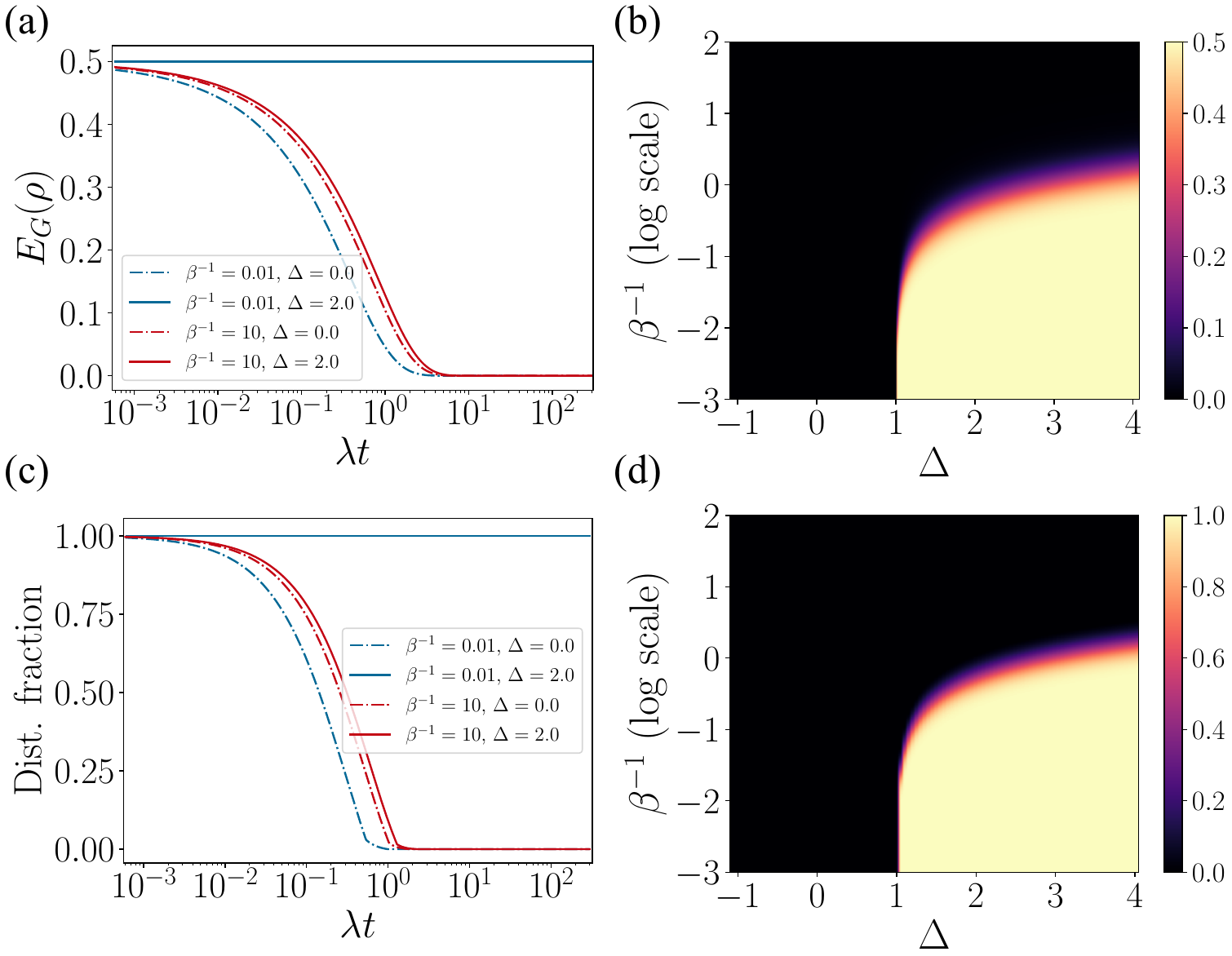}\\
\caption{{\color{black}{\emph{Dynamics of the generalized GHZ state.} (a,b) Geometric GME: (a) time evolution 
as a function of scaled time $\lambda t$ 
($\lambda = \gamma n$) for various 
temperatures and anisotropies, and (b) 
long-time behavior at $\lambda t = 100$. 
(c,d) GHZ distillation fraction: (c) time 
evolution and (d) long-time 
temperature--anisotropy dependence at 
$\lambda t = 100$. Parameters: 
$N/N_b \equiv f = 0.01$, $\gamma = 1$, 
$n = 10$. Both quantities show a sharp 
crossover at $\Delta = 1$: for $\Delta < 1$ 
the GHZ state rapidly decoheres, while 
for $\Delta > 1$ it remains robust up to 
an anisotropy-dependent threshold 
temperature.
 } } }
    \label{ghz_1}
\end{figure}

{\color{black} In Fig.~\ref{ghz_1}(a) and (c), we show the evolution of the Geometric GME and the distillation function as a function of a rescaled time $\lambda t$, where $\lambda = \gamma n$ and $n$ is the density per energy of the bath spins. We also show the variation of Geometric GME and distillation fraction at large times ($\lambda t =100 $) for various ranges of temperatures $\beta^{-1}$ and anistropy $\Delta$ in Fig.~\ref{ghz_1}(b) and (d). We notice that 
the dynamics of GME across the anisotropy landscape exhibits a clear distinction between the regimes $\Delta < 1$ and  $\Delta > 1$ in the presence of a thermal bath.  The region $\Delta >1$ seems to be favorable for GHZ type correlations. At large times the GHZ correlations survive till a threshold temperature $\beta_{c}^{-1}$ which depends on anisotropy $\Delta$. On the contrary, for $\Delta <1$, we find that the GHZ correlations decay rapidly at arbitrarily small temperatures. This decay can be attributed to the fact that for $\Delta <1$, creation of a single magnon is energetically more favorable than the zero magnon component $\ket{0}^{\otimes N}$ of the GHZ state. Similarly, due to the global spin-flip symmetry, a $N-1$ magnon state will be energetically more stable than the $\ket{1}^{\otimes N}$ state. Thus, coupling to a thermal bath with an infinitesimally small temperature will make the GHZ state lose coherence once $\Delta \leq 1$. Interestingly, this crossover point $\Delta=1$ also marks the Heisenberg isotropic point where the ground state changes from gapless to gapped~\cite{bethe}. However, it is important to note that the decoherence of GHZ is controlled by the spectral properties of the single magnon state above the polarized state and not by the critical properties. 
}


For {\color{black}$\Delta > 1$}, once the temperature exceeds the threshold, the onset of thermal fluctuations in the system leads to an inevitable loss of  GHZ correlations due to statistical mixing of eigenstates. The decay of GME  can be characterized in terms of an effective incoherent admixture induced by the bath. In the long-time limit, the dynamics generates a temperature-dependent mixing parameter 
{\color{black}$ p(\beta_c,\Delta-1) = x$}, 
set by the system energy scale {\color{black}$(\Delta-1)$}, assuming detectable GME persists only as long as this admixture remains below a threshold $x$. A characteristic temperature scale therefore becomes {\color{black}
$\beta_c^{-1} \sim 2(\Delta-1)/\ln[(1-x)/x]$}.
For some fixed $x$, this implies a logarithmic dependence of the GME retention temperature on anisotropy,
when plotted in semi-logarithmic axes, as evident from Fig.~\ref{ghz_1}(b) and (d). {\color{black} We estimate the threshold value to be $x \approx 0.007$ for both GME and distillation fraction by fitting the curves of $\beta^{-1}$ versus $\Delta$.} The temperature-anisotropy dependence of GME in our case exhibits a wedge-like structure reminiscent of the quantum-criticality observed near quantum phase transitions~\cite{sachdev}.


\par 


\emph{Generalized $W$ state:}
The generalized $W$ state $\sum_{i=1}^{N} \vert 0 \cdots 1_i 
\cdots 0 \rangle/\sqrt{N}$ represents a genuinely multipartite 
entangled state inequivalent to the GHZ class under LOCC, since 
it retains bipartite entanglement across all reduced 
subsystems~\cite{dur_2000}. In our setup, this state 
corresponds to a one-magnon Bethe eigenstate with zero total 
momentum~\cite{izyu,gaudin_2014}. The formal solution of the 
master equation initialized with the generalized $W$ state 
involves contributions from the zero-magnon eigenstate and 
two-magnon eigenstates, including both scattering and 
bound-state configurations.

\begin{figure}
\includegraphics[width=0.48\textwidth]{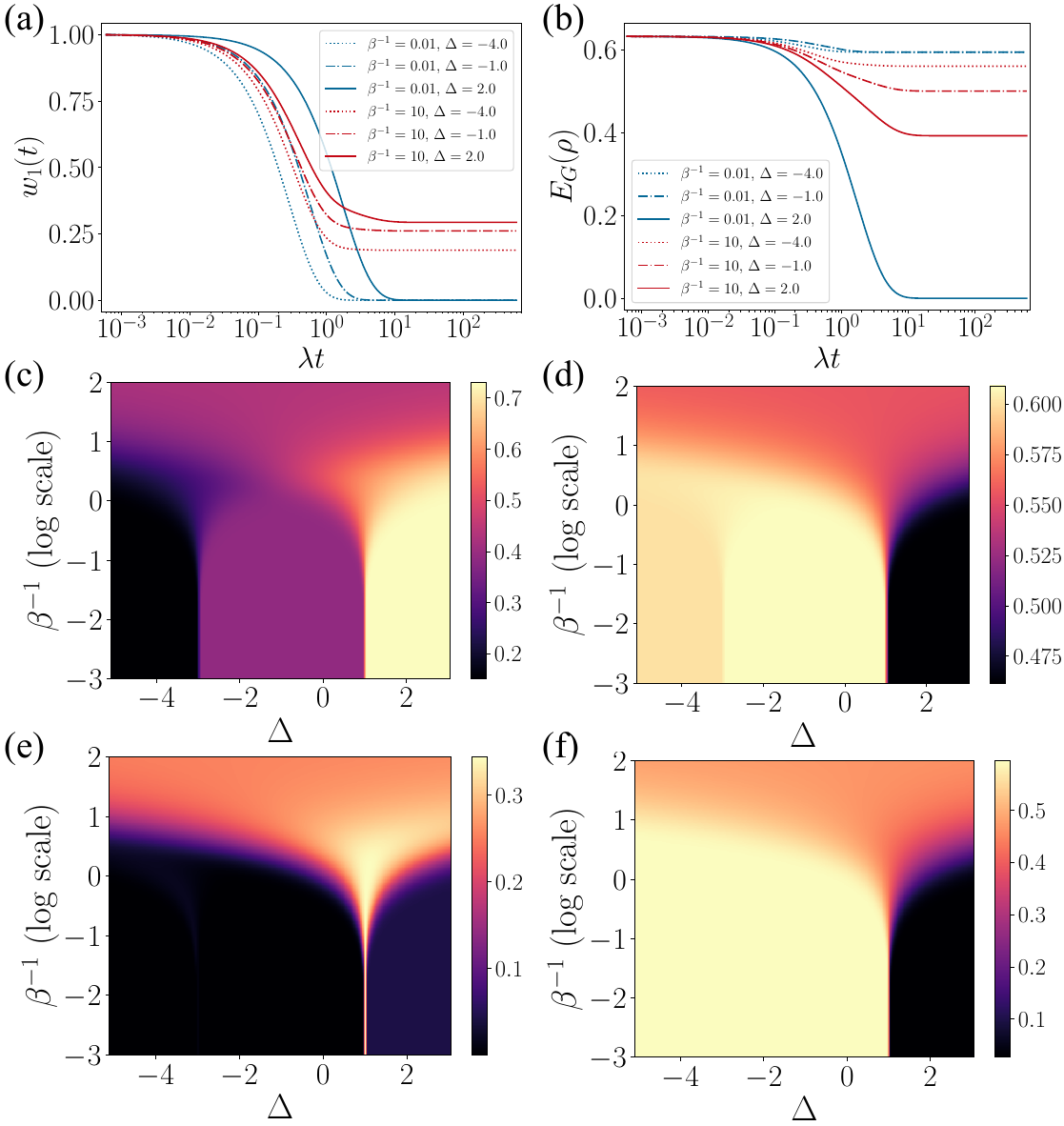}\\
\caption{{\color{black}{\emph{Dynamics of the generalized $W$ state.} 
(a,c,e) Fraction of the generalized $W$ 
state in the density matrix as a function 
of (a) scaled time $\lambda t$ 
($\lambda = \gamma n$), (c) temperature 
and anisotropy at $\lambda t = 10$, and 
(e) $\lambda t = 100$. (b,d,f) Geometric 
CME as a function of (b) scaled time, 
(d) temperature and anisotropy at 
$\lambda t = 10$, and (f) $\lambda t = 100$. 
Parameters: $N/N_b \equiv f = 0.01$, 
$\gamma = 1$, $n = 10$. The $W$-state 
fraction decays monotonically to $1/4$ 
at high temperatures for all $\Delta$, 
and to zero at low temperatures, except 
at $\Delta = 1$ where the zero-, one-, 
and two-magnon $q=0^+$ sectors are 
degenerate, resulting in an anomalously 
large asymptotic fraction. The CME 
vanishes for $\Delta > 1$ and remains 
nonzero for $\Delta \leq 1$, with 
$\Delta = 1$ marking the critical point 
between these regimes.}} }
    \label{w_1}
\end{figure}

In the thermodynamic limit, the two-magnon sector forms a 
continuum of states: the scattering sector, parametrized by a 
relative quasi-momentum $q \in (0,\pi)$, describes delocalized 
magnons, while the bound sector corresponds to localized 
spin-pair excitations~\cite{izyu,majumdar_pramana_1973} 
(see Section IA and IB in \cite{supplement}). As shown in Section II in \cite{supplement}, however, only the scattering 
states with quasi-momenta $q \to 0^+$ (translation-even) and 
$q \to \pi^-$ (translation-odd) contribute to the dynamics in 
the thermodynamic limit. The transition amplitudes associated 
with these states scale as $N/N_b$, whereas those associated 
with finite-$q$ scattering states and bound states scale as 
$1/(NN_b)$ and $1/N_b$, respectively, and are therefore 
suppressed in the thermodynamic limit. As a consequence, the 
closed-form time-dependent solution in the thermodynamic limit, 
expressed in the Bethe basis
$\{\ket{\Psi_0}, \ket{\Psi_1(q=0)}, \ket{\Psi_2(q=0^+)}, 
\ket{\Psi_2(q=\pi^-)}\}$, reads
\begin{eqnarray}
  \rho(t) = \mathrm{diag}\left(w_0,\, w_1, 
  w_2(0^+), w_2(\pi^-)\right).
 \label{eq: W_evolution}  
\end{eqnarray}
The time-dependent population terms $\mathbf{w} \equiv 
\left(w_0, w_1, w_2(0^+), w_2(\pi^-)\right)$ are obtained by 
solving the ME in Eq.~(\ref{eq:ME}) as
\begin{eqnarray}
\mathbf{w}(t) = \exp\!\left[\mathcal{M}(\beta, \Delta)t\right] 
\mathbf{w}(0),
\end{eqnarray}
where $\mathcal{M}(\beta, \Delta)$ is a Markov 
generator~\cite{gorini_jmp_1976,spohn_rmp_1980} encoding the 
dissipative transition rates between energy levels 
(see Section IIC in \cite{supplement} for the explicit expression of $\mathbf{w}(t)$). 
{\color{black}At low temperature, the structure of $\mathcal{M}(\beta,\Delta)$ 
permits only certain transitions: (i) for $\Delta > 1$, only 
$\ket{\Psi_1(q=0)} \to \ket{\Psi_0}$ is favored; (ii) for 
$-3 < \Delta < 1$, the transition $\ket{\Psi_1(q=0)} \to 
\ket{\Psi_2(q=0^+)}$ is favored; (iii) for $\Delta < -3 $, 
both the transitions transitions $\ket{\Psi_1(q=0)} \to \ket{\Psi_2(q=0^+)}$, and 
$\ket{\Psi_1(q=0)} \to \ket{\Psi_2(q=\pi^-)}$  are favored. As a consequence, the fraction 
of the $W$ state in the classical ensemble 
decays monotonically with time irrespective 
of anisotropy and temperature 
(Fig.~\ref{w_1}(a)), but the rate of decay 
depends strongly on anisotropy: it is 
fastest for $\Delta < -3$, intermediate 
for $-3 < \Delta < 1$, and slowest for 
$\Delta > 1$ (Fig.~\ref{w_1}(c)). At 
asymptotically long times, the fraction 
saturates to $1/4$ above the threshold 
temperature $\beta_c^{-1}$, since all four 
diagonal components of $\rho(t)$ are 
equally populated in the high-temperature 
limit, whereas it vanishes at low 
temperature. The exception is $\Delta = 1$, 
where the zero-, one-, and two-magnon 
$q=0^+$ sectors are mutually degenerate 
and no lower-energy state is accessible 
via the flip-flop bath coupling, resulting 
in an anomalously large asymptotic 
$W$-state fraction even at low temperatures 
(Fig.~\ref{w_1}(e)). }

The entanglement of the state in Eq.~\eqref{eq: W_evolution} 
is bounded by the convex roof criterion
$
E_G(\rho) \le w_1 E_G(\ket{\Psi_1(q=0)}) 
+ w_2(0^+) E_G(\ket{\Psi_2(q=0^+)})
+ w_2(\pi^-)E_G(\ket{\Psi_2(q=\pi^-)})$. The main challenge in computing the geometric GME in the 
thermodynamic limit stems from the non-extensivity of the 
measure. We show that for the one-magnon $W$ state and the 
two-magnon states $\ket{\Psi_2(q=0^+)}$ and 
$\ket{\Psi_2(q=\pi^-)}$, the geometric GME scales as 
$\mathcal{O}(N^{-1})$ with system size (refer to Section IVC in \cite{supplement} for details), 
rendering it unsuitable in the thermodynamic limit. We 
therefore compute the geometric measure of CME, defined 
with respect to the set of fully separable states in 
Eq.~\eqref{eq:convexity}~\cite{wei_jmp_2010}. In the 
thermodynamic limit, $E_G(\ket{\Psi_1(q=0)}) = 1 - e^{-1}$ 
and $E_G(\ket{\Psi_2(q=0^+)}) = E_G(\ket{\Psi_2(q=\pi^-)}) 
= 1 - 3e^{-2}$. We show (details in Section IVD in \cite{supplement}) that the convex roof 
bound is saturated exactly in the thermodynamic limit, so 
the CME of the state in Eq.~\eqref{eq: W_evolution} is given 
exactly by
\begin{eqnarray}
E_G(\rho) = \left(1 - \frac{1}{e}\right)w_1 
+ \left(1 - \frac{3}{e^2}\right)
\left(w_2(0^+) + w_2(\pi^-)\right).\nonumber\\
\end{eqnarray}

{\color{black}The CME shows strong anisotropy-dependence, 
particularly in the low-temperature regime: 
(i) for $\Delta > 1$, the CME vanishes at 
asymptotically long times as the system 
relaxes toward the unentangled ferromagnetic 
state $\ket{\Psi_0}$; (ii) for 
$-3 < \Delta < 1$, the long-time state is 
an incoherent mixture of $\ket{\Psi_0}$ and 
the entangled two-magnon state 
$\ket{\Psi_2(q=0^+)}$, preserving a 
nonzero CME; (iii) for $\Delta < -3$, the 
long-time state is an incoherent mixture 
of the entangled 
two-magnon states $\ket{\Psi_2(q = 0^+)}$  and $\ket{\Psi_2(q=\pi^-)}$, 
therefore preserving a nonzero CME. The CME 
thus vanishes asymptotically only for 
$\Delta > 1$ at low temperatures, while 
remaining nonzero for all $\Delta \leq 1$. 
Figures~\ref{w_1}(d) and (f) show these 
nontrivial crossovers at $\Delta = -3$ 
and $\Delta = 1$, separating the zero- and 
nonzero-CME regions at long times. Beyond 
the critical temperature, all eigenstates 
have finite transition probabilities, so 
the CME remains nonvanishing regardless 
of anisotropy; however, the one-excitation 
Dicke-type entanglement of the $W$ state 
is lost.}

{\color{black} 
For finite system sizes, local projective 
measurements on a symmetric two-excitation 
Dicke state can probabilistically collapse 
the remaining qubits into a $W$-type state~\cite{chiuri_prl_2012,witlef_prl_2009}. However no general LOCC protocol converts a 
Dicke state into an $N$-qubit $W$ state 
with finite distillation fraction in the 
large-$N$ limit~\cite{dur_2000,sauerwein_prx_2018}.}

\emph{Discussion and Conclusions:} Our model considers a broad class of experimentally relevant situations in which an interacting one-dimensional quantum magnet is embedded in an infinite spin bath, where anisotropy is tunable via chemical doping or applied strain~\cite{hase_prl_1993,takigawa_prl_1996,toskovic_np_2016,bounoua_prb_2017,witzel_prb_2006,prokofev_ropp_2000}. Trapped-ion platforms can provide controllable quantum simulators of XXZ chains  with anisotropy tunable via laser-mediated interactions~\cite{bloch_rmp_2008,friedenauer_np_2008,kim_n_2010}. {\color{black} In a realistic quasi-one-dimensional 
Cu-based spin chain with exchange energy 
$J \sim 5\,\mathrm{meV}$ 
($J/k_B \approx 58\,\mathrm{K}$)~\cite{
arai_prl_1996,zeisner_prb_2017}, the 
crossover temperature is estimated from 
the condition that the thermal admixture 
of one-magnon excitations reaches a 
threshold $x \approx 0.007$, giving 
$T_c \simeq 2(\Delta-1)J/[k_B\ln((1-x)/x)]$. 
For $\Delta \approx 1.3$--$2.0$, this 
yields $T_c \approx 8$--$20\,\mathrm{K}$, 
well within experimentally accessible 
low-temperature regimes, and robust 
against the precise value of $x$ owing 
to the logarithmic dependence of $T_c$ 
on the admixture fraction.}

{\color{black}Our central result is anisotropy-controlled 
crossovers in multipartite entanglement 
dynamics. The GHZ-type correlation, below a threshold, remain robust over time, above it they decay rapidly under thermal dissipation. For $W$-type states, collective multipartite entanglement shows strong anisotropy dependence, with two crossover points distinguishing low, intermediate, and high-entanglement regimes. In conclusion, we have 
analytically demonstrated that anisotropy 
in intra-system interactions plays a 
decisive role in stabilizing multipartite 
entanglement at low temperatures. Our 
results indicate that Hamiltonian 
engineering via anisotropy tuning can 
stabilize entanglement against thermal 
noise, providing a route to controlling 
intra-system interactions against colored 
noise from qubit-boson couplings. Through 
such control, it becomes possible to 
engineer entanglement-resilient subspaces 
without suppressing system-environment 
coupling, which is often technologically 
challenging in many quantum computing 
platforms.}

\section*{Acknowledgement} The authors are grateful to Sibasish Ghosh, Diptiman Sen, and V. Subrahmanyam for fruitful discussions. This work was supported by the Swiss National Science Foundation under Division II (Grant No. 200020-219400).

\bibliographystyle{unsrt}
\bibliography{manuscript.bib}

 \onecolumngrid

\clearpage
\input{Supplementary_arxiv.tex}

\end{document}

%% file: Supplementary_arxiv.tex
\linespread{1.0}

\begin{center}
    \textbf{Supplementary Material: Control of Multipartite Entanglement through Anisotropy against Thermal Noise}
\end{center}

\author{Samudra Sur}
\affiliation{Department of Quantum Matter Physics, University of Geneva, Quai Ernest-Ansermet 24, 1211 Geneva, Switzerland}

\author{Saikat Sur} \email{saikats@imsc.res.in}
\affiliation{Optics \& Quantum Information Group, The Institute of Mathematical Sciences, HBNI, CIT Campus, Taramani, Chennai 600113, India}

\maketitle

\onecolumngrid

\section{Bethe Ansatz solution for the anisotropic XXZ model}\label{AppA}

The Bethe ansatz wavefunction in the $l$-magnon sector is expressed as a superposition of plane waves~\cite{bethe,izyu}
\begin{eqnarray}
 |\Psi_l(\boldsymbol{q}) \rangle
 = \sum_{x_1,\ldots,x_l} \psi(\boldsymbol{q}\vert\boldsymbol{x})\,|\boldsymbol{x}\rangle,
\end{eqnarray}
with
\begin{equation}
\psi(\boldsymbol{q}|\boldsymbol{x})
= \sum_{\Pi} A_{\Pi}\,
\exp\!\left(i\sum_{j=1}^{l} q_{\Pi_j} x_j\right),
\end{equation}
where $q_1,\ldots,q_l$ are real or complex quasi-momenta of the magnons, $\Pi$ denotes permutations of these quasi-momenta, and the amplitudes $A_{\Pi}$ are fixed by two-body scattering phases.

The Bethe roots are obtained by solving the coupled set of $l$ Bethe equations
\begin{eqnarray}
N q_j &=& 2\pi I_j
 - \sum_{\substack{k=1 \\ k\neq j}}^{l} \Theta(q_j,q_k), \nonumber\\
\Theta(q_j,q_k) &=&
2\cot^{-1}\!\left(
\frac{\Delta \sin\frac{q_j-q_k}{2}}
{\cos\frac{q_j+q_k}{2}-\Delta\cos\frac{q_j-q_k}{2}}
\right),
\label{eq:bethe_eqn}
\end{eqnarray}
where $\Theta(q_j,q_k)$ is the two-body scattering phase related to the
S-matrix by $\mathcal{S}(q_j,q_k)=\exp[i\Theta(q_j,q_k)]$, and
$I_j\in\{0,1,\ldots,N-1\}$ are Bethe quantum numbers.
The corresponding energy eigenvalues are
\begin{equation}
\epsilon_l = -\frac{N\Delta}{2}
+ 2\sum_{j=1}^{l} (\Delta-\cos q_j).
\end{equation}

\begin{figure}[h]
\begin{center}
        \includegraphics[width=0.85\textwidth]{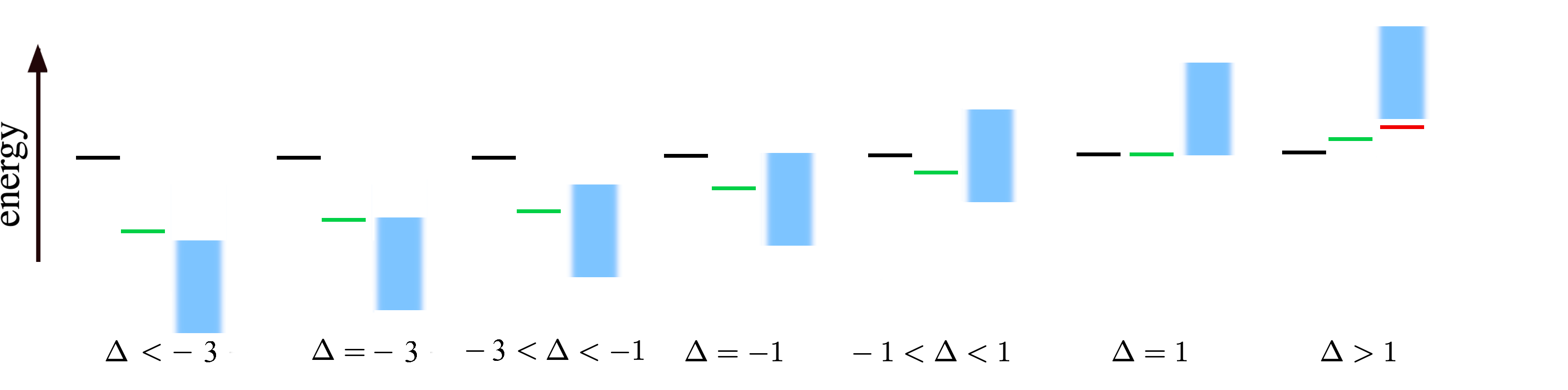}
\end{center}
\caption{{Zero-net-momentum energy eigenvalues of the XXZ model  in the thermodynamic limit $N \to \infty$, corresponding to the zero-magnon sector, the one-magnon sector, and the two-magnon scattering and bound-state sectors. These eigenvalues are represented by black, green, blue, and red lines, respectively.  }}
\label{fig:tmi}
\end{figure}

In this article, we consider only net momentum zero sector, i.e., $\sum_i q_i = 0$. The Bethe eigenvalues corresponding to the zero net-momentum are represented in Fig.~\ref{fig:tmi}.

\subsection{ One-magnon states}
Starting from the fully polarized ferromagnetic vacuum, a one-magnon excitation is created by flipping a single spin.
The one-magnon eigenstates are labeled by the momentum $p$ and have the wavefunction
\begin{equation}
\psi_q(x) = \frac{1}{\sqrt{N}} e^{iqx}, \qquad
q=\frac{2\pi I_1}{N},
\end{equation}
for a periodic chain, where $I_1=0,1,\ldots,N-1$.
The corresponding energy dispersion is
\begin{equation}
\epsilon_1(q) = -\frac{N\Delta}{2} + 2(\Delta-\cos q).
\end{equation}

\subsection{ Two-magnon scattering states}

For zero total momentum, $q_1+q_2=0$, we set $q_1=q$ and $q_2=-p$.
In the thermodynamic limit $N\rightarrow\infty$, the energy eigenvalues are
parametrized by a single continuous quantum number $q\in[0,\pi]$ as
\begin{eqnarray}
   \epsilon_2(q) = -\frac{N \Delta}{2} + 4 (\Delta - \cos q).
\end{eqnarray}

For $x_2>x_1$, the corresponding scattering eigenstates take the form
\begin{eqnarray}
   \psi_{q}(x_1,x_2) = A_q \left(e^{-iq(x_2-x_1)} + S(q,-q)\, e^{iq(x_2-x_1)}\right),
   \qquad q\in(0,\pi),
   \label{eq: sc_wf}
\end{eqnarray}
where $S(q,-q)=\exp[i\Theta(q,-q)]$ is the two-magnon scattering matrix.

The normalization factor in Eq.~\eqref{eq: sc_wf} is given by
\begin{eqnarray}
  A_q^{-2}
  &=& \sum^N_{x_1=1}\sum^{x_1+N-1}_{x_2=x_1+1} |\psi_q(x_1,x_2)|^2 \nonumber\\
  &=& N(N-1) + S(q,-q)\,\mathcal{I}_1(q,N) + S^*(q,-q)\,\mathcal{I}_2(q,N),
  \label{eq: norm}
\end{eqnarray}
where
\begin{eqnarray}
 \mathcal{I}_1(q,N) &=& \sum^N_{x_1=1}\sum^{x_1+N-1}_{x_2=x_1+1} e^{2iq(x_2-x_1)}
 = N e^{2iq}\frac{1-e^{2iq(N-1)}}{1-e^{2iq}}, \nonumber\\
 \mathcal{I}_2(q,N) &=& \sum^N_{x_1=1}\sum^{x_1+N-1}_{x_2=x_1+1} e^{-2iq(x_2-x_1)}
 = N e^{-2iq}\frac{1-e^{-2iq(N-1)}}{1-e^{-2iq}},
\end{eqnarray}
both of which scale as $\mathcal{O}(N)$, while the leading term scales as
$\mathcal{O}(N^2)$.

In the limits $q\to0^+$ and $q\to\pi^-$, the wavefunction
in Eq.~\eqref{eq: sc_wf} must be expanded to linear order in $q$.
This yields
\begin{eqnarray}
 \psi_{q=0^+}(x_1,x_2) &=& -2iA_q\,q\,(x_2-x_1), \nonumber\\
 \psi_{q=\pi^-}(x_1,x_2) &=& -2iA_q\,q\,(-1)^{x_2-x_1}(x_2-x_1).
\end{eqnarray}

The corresponding normalization factors are
\begin{eqnarray}
 A_q\big|_{q\to0^+} = A_q\big|_{q\to\pi^-}
 = \frac{1}{2qN\sqrt{\frac{(N-1)(2N-1)}{6}}}.
\end{eqnarray}

The normalized wavefunctions in these limits therefore become
\begin{eqnarray}
 \psi_{q}(x_1,x_2)\big|_{q=0^+} = \frac{-i\sqrt{3}}{N^2}(x_2-x_1), \qquad
 \psi_{q}(x_1,x_2)\big|_{q=\pi^-} = \frac{-i\sqrt{3}}{N^2}(-1)^{x_2-x_1}(x_2-x_1).
 \label{eq: sc_wf_end}
\end{eqnarray}

\subsection{ Two-magnon bound states}

The two-magnon scattering matrix for the XXZ chain can be written as~\cite{izyu,majumdar_pramana_1973}
\begin{eqnarray}
    \mathcal{S}(q_1,q_2)
    = \exp[i\Theta(q_1,q_2)]
    = -\frac{\cos\frac{Q}{2}-\Delta e^{iq}}
    {\cos\frac{Q}{2}-\Delta e^{-iq}},
\end{eqnarray}
where $Q=q_1+q_2$ is the total momentum and
$q=(q_1-q_2)/2$ is the relative momentum.

Bound states correspond to poles of the scattering matrix, which occur when
the denominator vanishes,
\begin{eqnarray}
    \cos\frac{Q}{2}-\Delta e^{-iq}=0.
\end{eqnarray}

For zero total momentum $Q=0$, we set
$p_1=i\kappa/2$ and $q_2=-i\kappa/2$ with $\kappa>0$.
The pole condition then yields
\begin{eqnarray}
    \Delta e^{-\kappa/2}=1,
\end{eqnarray}
leading to
\begin{eqnarray}
    \kappa = 2\ln\Delta.
\end{eqnarray}
Thus, a two-magnon bound state exists only for $\Delta>1$.

The corresponding energy eigenvalue is
\begin{eqnarray}
    \epsilon_2
    &=& -\frac{N\Delta}{2}
    +4\left(\Delta-\cos\frac{q_1-q_2}{2}\right) \nonumber\\
    &=& -\frac{N\Delta}{2}
    +4\left(\Delta-\cosh\frac{\kappa}{2}\right) \nonumber\\
    &=& -\frac{N\Delta}{2}
    +2\left(\Delta-\frac{1}{\Delta}\right).
\end{eqnarray}

The corresponding bound-state wavefunction for $x_2>x_1$ is
\begin{eqnarray}
    \psi_{\text{bound}}(x_1,x_2)
    = A\, e^{-\ln\Delta\,(x_2-x_1)}.
\label{eq: bound_wf}
\end{eqnarray}

The normalization condition
\begin{eqnarray}
    \sum_{x_1=1}^N\sum_{x_2=x_1+1}^{x_1+N-1}
    |\psi_{\text{bound}}(x_1,x_2)|^2 = 1
\end{eqnarray}
gives
\begin{eqnarray}
    A=\sqrt{\frac{\Delta^2-1}{N}}.
\end{eqnarray}

\section{Derivation and Solution of the Master Equation}
\label{AppB}

Under weak system--bath coupling, the reduced dynamics of the system
is described by the second-order time-convolutionless (TCL) master equation.
In the interaction picture, the reduced density matrix $\rho_S(t)$ obeys~\cite{petruccione}
\begin{subequations}
\begin{align}
\dot{\rho}_S(t)
= -\int_0^t d\tau\, \Sigma^{(2)}(t,\tau)\,\rho_S(t),
\label{eq:ME}
\end{align}
where the second-order self-energy superoperator is
\begin{align}
\Sigma^{(2)}(t,\tau)
= \mathrm{Tr}_B
\Big[
H_{SB}(t),
\big[
H_{SB}(t-\tau),
(\cdot)\rho_B
\big]
\Big].
\end{align}
\end{subequations}

Here
\[
H_{SB}(t)
= e^{i(H_S+H_B)t} H_{SB} e^{-i(H_S+H_B)t}
\]
is the system--bath interaction Hamiltonian in the interaction picture,
and $\rho_B$ is the equilibrium density matrix of the bath.

We assume a bilinear coupling of the form
\[
H_{SB}
= S_1 \otimes B_1 + S_2 \otimes B_2,
\]
with $S_1=\sum_i\sigma_i^-$ and $S_2=\sum_i\sigma_i^+$.
Under the Born approximation and assuming an initially factorized state,
the master equation can be written as
\begin{eqnarray}
\dot{\rho}_S(t)
&=&
S_1(t)\rho_S(t)W_2(t)
+ S_2(t)\rho_S(t)W_1(t)
- W_1(t)S_2(t)\rho_S(t)
- W_2(t)S_1(t)\rho_S(t)
+ \mathrm{H.c.},
\label{eq:ME1}
\end{eqnarray}
where, in the Markov limit, the operators
\begin{eqnarray}
W_1(t) &=& \int_0^\infty ds\, S_1(t-s)\,\Phi_{21}(s), \nonumber\\
W_2(t) &=& \int_0^\infty ds\, S_2(t-s)\,\Phi_{12}(s),
\label{eq:W}
\end{eqnarray}
are defined in terms of the bath correlation functions.

The system operators admit a block-diagonal representation in the Bethe eigenbasis,
\begin{eqnarray}
S_1(t)
&=&
\bigoplus_l
\sum_{\boldsymbol{p}_l,\boldsymbol{p}'_{l+1}}
e^{-it(\epsilon_l-\epsilon_{l+1})}
\langle\Psi_{l+1}|\sum_i\sigma_i^-|\Psi_l\rangle
|\Psi_{l+1}\rangle\langle\Psi_l|,
\nonumber\\
S_2(t)
&=&
\bigoplus_l
\sum_{\boldsymbol{p}'_l,\boldsymbol{p}_{l+1}}
e^{-it(\epsilon_{l+1}-\epsilon_l)}
\langle\Psi_l|\sum_i\sigma_i^+|\Psi_{l+1}\rangle
|\Psi_l\rangle\langle\Psi_{l+1}|.
\label{eq:S}
\end{eqnarray}

\begin{eqnarray}
&&\Phi_{12}(\tau) = \sum_{k_1,k_2 } g_{k_1}  g_{k_2} (C_{k_1} + i S_{k_1}) \langle \sigma^-_{k_1} \sigma^+_{k_2}\rangle =  \sum_{k } g^2_{k}  (C_{k} + i S_{k}) \langle \sigma^-_{k} \sigma^+_{k}\rangle =   \gamma \big\langle e^{2i\omega \tau} p(-\omega) \big\rangle, \nonumber\\
&&\Phi_{21}(\tau) = \sum_{k_1,k_2 } g_{k_1}  g_{k_2} (C_{k_1} - i S_{k_1}) \langle \sigma^+_{k_1} \sigma^-_{k_2}\rangle =  \sum_{k } g^2_{k}  (C_{k} - i S_{k}) \langle \sigma^+_{k} \sigma^-_{k}\rangle = \gamma \big\langle e^{-2i\omega \tau} p(\omega) \big\rangle, \nonumber\\
\end{eqnarray}

with $\gamma = \sum_{k} g^2_k$, is a finite positive number dependent on the nature of interaction between the bath spins and the geometrical/spatial arrangements of the spins; $n(\omega)$ is the ensemble spectral density normalized to the total number of spins; $\int d\omega~n(\omega) = N_{bath}$; $p(\omega) = \exp(-\beta \omega)/(2 \cosh(\beta \omega))$ and $\big\langle f(\omega) \big\rangle = \int d\omega~n(\omega) f(\omega)/ \int d\omega~n(\omega)$.

Spin systems often possess long-range exchange interactions scaling with distance $r$ as $g_k = C/r^\nu$ with the resonant dipole-dipole interaction, $\nu = 3$, which is the most common case~\cite{nick}. Let us assume that the bath spins are uniformly scattered in the space with density $\sigma_0$, $\gamma$ can be computed as follows. In $d$-dimensional space, the number of particles inside two concentric spheres of radius $r$ and $r+dr$ is $\sigma_0A_d r^{d-1}dr$, where $A_d = \frac{2\pi^{d/2}}{\Gamma(\frac{d}{2})}$ is the surface area of a $d$-dimensional sphere of unit radius. Therefore,

\begin{eqnarray}
 \gamma = \sum_k g_k^2 &=& C^2\sigma_0 A_d \int^\infty_a dr~\frac{r^{d-1}}{r^{2\nu}} =  C^2\sigma_0 \frac{2\pi^{d/2}}{\Gamma(\frac{d}{2})}\frac{a^{d-2\nu}}{2\nu - d},~\text{~~assuming~}2\nu > d. 
\end{eqnarray}
with $a$ being the ultraviolet cutoff length.

Starting from Eqs.~\eqref{eq:S} and \eqref{eq:W}, the dissipative operators entering the master equation can be expressed explicitly in the Bethe eigenbasis as
\begin{align}
W_1(t) &= \frac{\gamma}{N_{\mathrm{bath}}}
\bigoplus_{l=0}^{N}
\sum_{\boldsymbol{p}_l,\boldsymbol{p}'_{l+1}}
e^{-it(\epsilon_l(\boldsymbol{p}_l)-\epsilon_{l+1}(\boldsymbol{p}'_{l+1}))}
\langle \Psi_{l+1}(\boldsymbol{p}'_{l+1}) |
\sum_i \sigma_i^- |
\Psi_l(\boldsymbol{p}_l) \rangle \times
|\Psi_{l+1}(\boldsymbol{p}'_{l+1})\rangle
\langle \Psi_l(\boldsymbol{p}_l)|
\int d\omega\, p(\omega)\,n(\omega)\,\mathcal{I}_1(\omega),
\\[1ex]
W_2(t) &= \frac{\gamma}{N_{\mathrm{bath}}}
\bigoplus_{l=0}^{N}
\sum_{\boldsymbol{p}'_l,\boldsymbol{p}_{l+1}}
e^{-it(\epsilon_{l+1}(\boldsymbol{p}_{l+1})-\epsilon_l(\boldsymbol{p}'_l))}
\langle \Psi_l(\boldsymbol{p}'_l) |
\sum_i \sigma_i^+ |
\Psi_{l+1}(\boldsymbol{p}_{l+1}) \rangle
\times
|\Psi_l(\boldsymbol{p}'_l)\rangle
\langle \Psi_{l+1}(\boldsymbol{p}_{l+1})|
\int d\omega\, p(-\omega)\,n(\omega)\,\mathcal{I}_2(\omega).
\label{eq:W_SI}
\end{align}

Including the Lamb-shift terms, the frequency integrals appearing in Eq.~\eqref{eq:W_SI} evaluate to
\begin{align}
\int_{-\infty}^{\infty} d\omega\, p(\omega)n(\omega)\mathcal{I}_1(\omega)
&=
\frac{\pi}{2}
p\!\left(\frac{\epsilon_l-\epsilon_{l+1}}{2}\right)
n\!\left(\frac{\epsilon_l-\epsilon_{l+1}}{2}\right)
\nonumber\\
&\quad
- i \int_{-\infty}^{\infty} d\omega\,
p(\omega)n(\omega)\,
\mathcal{P}\!\left(
\frac{1}{2\omega-(\epsilon_l-\epsilon_{l+1})}
\right),
\\[1ex]
\int_{-\infty}^{\infty} d\omega\, p(-\omega)n(\omega)\mathcal{I}_2(\omega)
&=
\frac{\pi}{2}
p\!\left(\frac{\epsilon_{l+1}-\epsilon_l}{2}\right)
n\!\left(\frac{-\epsilon_{l+1}+\epsilon_l}{2}\right)
\nonumber\\
&\quad
+ i \int_{-\infty}^{\infty} d\omega\,
p(-\omega)n(\omega)\,
\mathcal{P}\!\left(
\frac{1}{2\omega+(\epsilon_{l+1}-\epsilon_l)}
\right),
\end{align}
where $\mathcal{P}$ denotes the Cauchy principal value.

From Eq.~\eqref{eq:S} and  Eq.~\eqref{eq:W_SI}, the explicit form of the operators are recast as
\begin{eqnarray}
  S^{01}_1(t) &=&  \sqrt{N} ~e^{-2it(1-\Delta )}, \nonumber\\
  S^{10}_2(t) &=&  \sqrt{N} ~e^{2it(1-\Delta)}, \nonumber\\ 
  W^{01}_1(t) &=&  \frac{\gamma \sqrt{N}}{N_{b}}  ~e^{-2it(1-\Delta )} \left[\frac{\pi}{2} p(1-\Delta ) n(1-\Delta ) - i\int^{\infty}_{-\infty} d\omega~ p(\omega) n(\omega) \mathcal{P}\left( \frac{1}{ 2 \omega  + 2(\Delta - 1)}\right) \right], \nonumber\\
  W^{10}_2(t) &=&   \frac{\gamma \sqrt{N}}{N_{b}}  ~e^{2it(1-\Delta)} \left[\frac{\pi}{2} p(\Delta - 1 ) n(1-\Delta) + i\int^{\infty}_{-\infty} d\omega~ p(-\omega) n(\omega) \mathcal{P}\left( \frac{1}{ 2 \omega  + 2(\Delta - 1)}\right) \right]. \nonumber\\
  \end{eqnarray}

\subsection{ Reduced dynamics of zero magnon sector}

Restricting to the zero- and one-magnon sectors, the reduced density matrix in the Bethe eigenbasis reads
\begin{align}
\rho(t) &=
\rho_{00}(t)|\Psi_0\rangle\langle\Psi_0|
+ \rho_{01}(t)|\Psi_0\rangle\langle\Psi_1(\boldsymbol{p}=0)|
\nonumber\\
&\quad
+ \rho_{01}^*(t)|\Psi_1(\boldsymbol{p}=0)\rangle\langle\Psi_0|
+ \rho_{11}(t)|\Psi_1(\boldsymbol{p}=0)\rangle
\langle\Psi_1(\boldsymbol{p}=0)| + \cdots .
\end{align}

In this truncated basis, the operators $S_1(t)$, $S_2(t)$, $W_1(t)$, and $W_2(t)$ reduce to
\begin{align}
S_1(t) &= \begin{pmatrix}
0 & 0\\
S^{01}_1(t) & 0
\end{pmatrix},\qquad
S_2(t) = \begin{pmatrix}
0 & S^{10}_2(t)\\
0 & 0
\end{pmatrix},\nonumber\\
W_1(t) &= \begin{pmatrix}
0 & 0\\
W^{01}_1(t) & 0
\end{pmatrix},\qquad
W_2(t) = \begin{pmatrix}
0 & W^{10}_2(t)\\
0 & 0
\end{pmatrix}.
\end{align}

The master equation Eq.~\eqref{eq:ME1} then simplifies to
\begin{align}
\begin{pmatrix}
\dot{\rho}_{00} & \dot{\rho}_{01}\\
\dot{\rho}_{01}^* & \dot{\rho}_{11}
\end{pmatrix}
=
\begin{pmatrix}
S^{10}_2\rho_{11}W^{01}_1 - W^{10}_2\rho_{00}S^{01}_1 + \mathrm{h.c.}
&
W^{10}_2\rho_{01}S^{01}_1 + \mathrm{h.c.}
\\
W^{01}_1\rho_{01}^*S^{10}_2 + \mathrm{h.c.}
&
S^{01}_1\rho_{00}W^{10}_2 - W^{01}_1\rho_{11}S^{10}_2 + \mathrm{h.c.}
\end{pmatrix}.
\end{align}

The evolution of the coherence $\rho_{01}(t)$ is governed by
\begin{align}
\dot{\rho}_{01}(t)
=
W^{10}_2(t)\rho_{01}(t)S^{01}_1(t)
+
W^{*10}_2(t)\rho^*_{01}(t)S^{*01}_1(t).
\end{align}
Separating real and imaginary parts and imposing the initial conditions
$\Re[\rho_{01}(0)]=\Im[\rho_{01}(0)]=0$, we obtain
\begin{align}
\Re[\rho_{01}(t)] = 0,
\qquad
\Im[\rho_{01}(t)] = 0.
\end{align}
Thus, coherences are not dynamically generated in this truncated subspace.

The population dynamics are given by
\begin{align}
\dot{\rho}_{00}(t) &=
S^{10}_2\rho_{11}W^{01}_1 - W^{10}_2\rho_{00}S^{01}_1 + \mathrm{h.c.},\\
\dot{\rho}_{11}(t) &=S^{01}_1\rho_{00}W^{10}_2 - W^{01}_1\rho_{11}S^{10}_2 + \mathrm{h.c.}.
\end{align}
Using $\rho_{00}(t)+\rho_{11}(t)=1$, we obtain a closed equation for $\rho_{11}(t)$:
\begin{align}
&&\dot{\rho}_{11}(t)
=
\left(S^{01}_1W^{10}_2 + S^{*01}_1W^{*10}_2\right)
-
\rho_{11}(t)
\left(
S^{01}_1W^{10}_2
+ W^{01}_1S^{10}_2
+ S^{*01}_1W^{*10}_2
+ W^{*01}_1S^{*10}_2
\right).\\
&&\implies \rho_{11}(t)= 1 - \frac{\Re[S^{01}_1W^{10}_2]}{\Re[S^{10}_2W^{01}_1
+ W^{10}_2S^{01}_1]}(1 - e^{-2t(\Re[S^{10}_2W^{01}_1
+ W^{10}_2S^{01}_1])}).
\end{align}

\subsection{ Decay of extremal coherences: $\boldsymbol{|\Psi_0\rangle\langle\Psi_N|}$ sector}

We consider an initial coherence between the fully polarized ferromagnetic states,
$\rho(0)=|\Psi_0\rangle\langle\Psi_N|$.
Under the system–bath dynamics generated by Eq.~\eqref{eq:ME1}, the reduced density matrix at time $t$ is confined to the subspace
$\{|\Psi_0\rangle,|\Psi_1\rangle,|\Psi_{N-1}\rangle,|\Psi_N\rangle\}$ and can be written as
\begin{align}
\rho(t) &=
\rho_{0N}(t)|\Psi_0\rangle\langle\Psi_N|
+ \rho_{0,N-1}(t)|\Psi_0\rangle\langle\Psi_{N-1}|
+ \rho_{1,N-1}(t)|\Psi_1\rangle\langle\Psi_{N-1}|
+ \rho_{1N}(t)|\Psi_1\rangle\langle\Psi_N|.
\end{align}

In the ordered Bethe basis
$\{|\Psi_0\rangle,|\Psi_1\rangle,|\Psi_{N-1}\rangle,|\Psi_N\rangle\}$,
the operators $S_{1,2}(t)$ and $W_{1,2}(t)$ take the block form
\begin{align}
S_1(t) &=
\begin{pmatrix}
0 & 0 & 0 & 0\\
S^{01}_1(t) & 0 & 0 & 0\\
0 & 0 & 0 & 0\\
0 & 0 & S^{N-1,N}_1(t) & 0
\end{pmatrix},
\qquad
S_2(t) =
\begin{pmatrix}
0 & S^{10}_2(t) & 0 & 0\\
0 & 0 & 0 & 0\\
0 & 0 & 0 & S^{N,N-1}_2(t)\\
0 & 0 & 0 & 0
\end{pmatrix},\nonumber\\[1ex]
W_1(t) &=
\begin{pmatrix}
0 & 0 & 0 & 0\\
W^{01}_1(t) & 0 & 0 & 0\\
0 & 0 & 0 & 0\\
0 & 0 & W^{N-1,N}_1(t) & 0
\end{pmatrix},
\qquad
W_2(t) =
\begin{pmatrix}
0 & W^{10}_2(t) & 0 & 0\\
0 & 0 & 0 & 0\\
0 & 0 & 0 & W^{N,N-1}_2(t)\\
0 & 0 & 0 & 0
\end{pmatrix}.
\end{align}

Projecting Eq.~\eqref{eq:ME1} onto this subspace yields
\begin{align}
\begin{pmatrix}
0 & 0 & \dot{\rho}_{0,N-1} & \dot{\rho}_{0N}\\
0 & 0 & \dot{\rho}_{1,N-1} & \dot{\rho}_{1N}\\
\dot{\rho}^*_{0,N-1} & \dot{\rho}^*_{1,N-1} & 0 & 0\\
\dot{\rho}^*_{0N} & \dot{\rho}^*_{1N} & 0 & 0
\end{pmatrix}
=
\begin{pmatrix}
0 & 0 & A_{0,N-1} & A_{0N}\\
0 & 0 & A_{1,N-1} & A_{1N}\\
A_{N-1,0} & A_{N-1,1} & 0 & 0\\
A_{N,0} & A_{N,1} & 0 & 0
\end{pmatrix}
+ \mathrm{h.c.}
\end{align}
with
\begin{align}
A_{0,N-1} &= W^{N,N-1}_2\rho_{1N}S^{01}_1
           - W^{01}_1\rho_{0,N-1}S^{10}_2, \nonumber\\
A_{0N} &= - W^{01}_1\rho_{0N}S^{10}_2, \nonumber\\
A_{1,N-1} &= - W^{10}_2\rho_{1,N-1}S^{01}_1, \nonumber\\
A_{1N} &= W^{N,N-1}_2\rho_{0,N-1}S^{01}_1
         - W^{01}_1\rho_{1N}S^{10}_2,
\end{align}
and Hermitian conjugates defining the remaining entries.

Using spin-flip symmetry of the XXZ Hamiltonian,
\begin{align}
S^{01}_1 &= S^{N,N-1}_2, &
S^{10}_2 &= S^{N-1,N}_1, \nonumber\\
W^{01}_1 &= W^{N,N-1}_2, &
W^{10}_2 &= W^{N-1,N}_1,
\end{align}
the equations for the extremal coherences $(0,N)$ and $(1,N-1)$ decouple.

From the $(1,4)$ and $(4,1)$ components we obtain
\begin{align}
\dot{\rho}_{0N}
&= -W^{10}_2 S^{01}_1 \rho_{0N}
   - W^{*10}_2 S^{*01}_1 \rho^*_{0N}, \nonumber\\
\dot{\rho}^*_{0N}
&= -W^{10}_2 S^{01}_1 \rho^*_{0N}
   - W^{*10}_2 S^{*01}_1 \rho_{0N}.
\end{align}
Separating real and imaginary parts,
\begin{align}
\Re[\dot{\rho}_{0N}] &= -2\Re\!\left(W^{10}_2 S^{01}_1\right)\Re[\rho_{0N}],\nonumber\\
\Im[\dot{\rho}_{0N}] &= 0,
\end{align}
leading to the exponential decay
\begin{align}
\Re[\rho_{0N}(t)]
= \Re[\rho_{0N}(0)]
\exp\!\left[-2t\,\Re\!\left(W^{10}_2 S^{01}_1\right)\right].
\end{align}

Analogously, from the $(2,3)$ and $(3,2)$ components,
\begin{align}
\Re[\rho_{1,N-1}(t)]
= \Re[\rho_{1,N-1}(0)]
\exp\!\left[-2t\,\Re\!\left(W^{01}_1 S^{10}_2\right)\right],
\qquad
\Im[\rho_{1,N-1}(t)] = 0.
\end{align}

\subsection{ Reduced Dynamics of the generalized $W$ state}

From Eqs.~\eqref{eq:S} and \eqref{eq:W_SI}, the explicit forms of the transition
operators coupling the one-magnon and two-magnon sectors can be written as
\begin{eqnarray}
  S^{12}_1(q,t) &=&  \Omega_q^{*}\,
  e^{-2it(2\cos q - 1 - 2\Delta)}, \nonumber\\
  S^{21}_2(q,t) &=&  \Omega_q\,
  e^{2it(2\cos q - 1 - 2\Delta)}, \nonumber\\
  W^{12}_1(q,t) &=&
  \frac{\gamma\,\Omega_q^{*}}{N_b}
  e^{-2it(2\cos q - 1 - 2\Delta)}
  \Bigg[
  \frac{\pi}{2}\,
  p(2\cos q - 1 - \Delta)\,
  n(2\cos q - 1 - \Delta)
  \nonumber\\
  &&\hspace{2.5cm}
  -\, i \int_{-\infty}^{\infty} d\omega\,
  p(\omega)n(\omega)\,
  \mathcal{P}\!\left(
  \frac{1}{2\omega + 2\Delta + 2 - 4\cos q}
  \right)
  \Bigg], \nonumber\\
  W^{21}_2(q,t) &=&
  \frac{\gamma\,\Omega_q}{N_b}
  e^{2it(2\cos q - 1 - 2\Delta)}
  \Bigg[
  \frac{\pi}{2}\,
  p(1+\Delta-2\cos q)\,
  n(2\cos q - 1 - \Delta)
  \nonumber\\
  &&\hspace{2.5cm}
  +\, i \int_{-\infty}^{\infty} d\omega\,
  p(-\omega)n(\omega)\,
  \mathcal{P}\!\left(
  \frac{1}{2\omega + 2\Delta + 2 - 4\cos q}
  \right)
  \Bigg]. \nonumber
\end{eqnarray}
Here $\mathcal{P}$ denotes the Cauchy principal value.

The transition amplitude $\Omega_q$ is obtained from the two-magnon scattering
and bound-state wavefunctions,
Eqs.~\eqref{eq: sc_wf}, \eqref{eq: sc_wf_end}, and \eqref{eq: bound_wf},
\begin{equation}
\Omega_q
=
\langle \Psi_1 |
\sum_i \sigma_i^{+}
| \Psi_2(q) \rangle
=
\frac{2}{\sqrt{N}}
\sum_{x_1=1}^{N}
\sum_{x_2=x_1+1}^{x_1+N-1}
\psi_q(x_1,x_2)
=
\mathcal{O}\!\left(\frac{1}{\sqrt{N}}\right),
\end{equation}
for generic momenta $q$.

However, near the band edges the overlap is enhanced,
\begin{eqnarray}
\lim_{q\to 0^{+}} \Omega_q
&=&
-\,i\sqrt{3N}, \nonumber\\
\lim_{q\to \pi^{-}} \Omega_q
&=&
+\,i\sqrt{3N},
\end{eqnarray}
indicating a collective enhancement of the transition matrix element.
For the two-magnon bound state,
\begin{equation}
\Omega_{\mathrm{bound}}
=
\langle \Psi_1 |
\sum_i \sigma_i^{+}
| \Psi_2(\mathrm{bound}) \rangle
=
2\sqrt{\frac{\Delta+1}{\Delta-1}},
\end{equation}
which remains $\mathcal{O}(1)$ in the thermodynamic limit.

Retaining only the dominant modes
$\{|\Psi_0\rangle, |\Psi_1(q=0)\rangle,
|\Psi_2(q=0^{+})\rangle, |\Psi_2(q=\pi^{-})\rangle\}$,
the master equation~\eqref{eq:ME1} reduces to
\begin{equation}
\dot{\rho}(t)
=
\mathcal{M}(\beta,\Delta)\,\rho(t).
\end{equation}
Within the secular approximation, the density matrix remains diagonal in this
basis,
\begin{equation}
\rho(t)
=
\begin{pmatrix}
w_0 & 0 & 0 & 0\\
0 & w_1 & 0 & 0\\
0 & 0 & w_2(0^{+}) & 0\\
0 & 0 & 0 & w_2(\pi^{-})
\end{pmatrix}.
\end{equation}

The Markov generator takes the form
\begin{eqnarray}
\mathcal{M}(\beta,\Delta)
=
\begin{pmatrix}
- S^{01}_1 W^{10}_2
& S^{10}_2 W^{01}_1
& 0 & 0\\
S^{01}_1 W^{10}_2
& -\!\!\sum_{q=0^{+},\pi^{-}}\! S^{12}_1(q)W^{21}_2(q) - S^{10}_2 W^{01}_1
& S^{21}_2(0^{+})W^{12}_1(0^{+})
& S^{21}_2(\pi^{-})W^{12}_1(\pi^{-})\\
0
& S^{12}_1(0^{+})W^{21}_2(0^{+})
& -S^{21}_2(0^{+})W^{12}_1(0^{+})
& 0\\
0
& S^{12}_1(\pi^{-})W^{21}_2(\pi^{-})
& 0
& -S^{21}_2(\pi^{-})W^{12}_1(\pi^{-})
\end{pmatrix}.
\end{eqnarray}
By construction, $\sum_i \dot w_i = 0$, ensuring probability conservation.

One of the eigenvalues of the matrix $\mathcal{M}(\beta,\Delta)$ is zero and the rest three eigenvalues  are determined from the following cubic equation 

\begin{eqnarray}
&&  \lambda^3
+
(a_1+a_2+a_3+b_1+b_2+b_3)\lambda^2
+
\big(
a_1a_2+a_1a_3+a_2a_3
+ a_1(b_2+b_3)
+ a_2(b_1+b_3)
+ a_3(b_1+b_2)
\big)\lambda
 \nonumber\\
&& + \big(
a_1a_2b_3
+
a_1a_3b_2
+
a_2a_3b_1
\big)= 0,   
\end{eqnarray}
and the corresponding eigenvectors are of the form
\begin{eqnarray}
\mathbf{v}^{(\lambda)}
=
\begin{pmatrix}
\dfrac{b_1}{a_1+\lambda} \\[8pt]
1 \\[8pt]
\dfrac{b_2}{a_2+\lambda} \\[8pt]
\dfrac{b_3}{a_3+\lambda}
\end{pmatrix},
\end{eqnarray}

where

\begin{eqnarray}
 a_1 &=& 2 \Re{S^{01}_1 W^{10}_2}, \nonumber\\ 
a_2 &=& 2\Re{S^{21}_2(0^+) W^{12}_1(0^+)}, \nonumber\\
 a_3 &=& 2\Re{S^{21}_2(\pi^-) W^{12}_1(\pi^-)}, \nonumber\\
  b_1 &=& 2\Re{S^{10}_2 W^{01}}_1, \nonumber\\ 
b_2 &=& 2\Re{S^{12}_1(0^+) W^{21}_2(0^+)}, \nonumber\\
 b_3 &=& 2\Re{S^{12}_1(\pi^-) W^{21}_2(\pi^-)}. \nonumber\\
\end{eqnarray}

The time-dependent population vector $
\mathbf{w}(t)
\equiv
\Big(
w_0(t),
w_1(t),
w_2(0^+)(t),
w_2(\pi^-)(t)
\Big)$ 
can therefore be written as
\begin{equation}
\mathbf{w}(t)
=
\sum_{\lambda}
e^{\lambda t}
c_{\lambda}
\mathbf{v}^{(\lambda)},
\qquad
\text{where}
\qquad
\mathbf{w}(0)
=
\sum_{\lambda}
c_{\lambda}
\mathbf{v}^{(\lambda)} .
\end{equation}

\section{Distillation protocol for the generalized GHZ state}\label{AppC}
\subsection{GHZ distillation via bilateral CNOT and postselection}
We briefly summarize the standard multipartite GHZ distillation 
protocol~\cite{murao_pra_1998}. A general $N$-qubit state is written as
\begin{equation}
\rho = \sum_{x,y \in \{0,1\}^N} \rho_{x,y}\,|x\rangle\langle y|.
\end{equation}

\paragraph*{Step 1: Bilateral CNOT operation.}
Two identical copies of the state are prepared, labeled as control (C) 
and target (T). Each of the $N$ parties applies a local CNOT gate, with 
the control qubit acting on the corresponding target qubit.
The global unitary thus maps
\begin{equation}
|x\rangle_C |u\rangle_T \;\longmapsto\; |x\rangle_C |u \oplus x\rangle_T,
\label{eq:CNOT}
\end{equation}
where $\oplus$ denotes bitwise XOR. This operation correlates phase 
information in the control copy with bit-flip information in the target 
copy.

\paragraph*{Step 2: Collective parity measurement.}
After the CNOT layer, the target register undergoes a collective even-parity 
measurement, projecting onto the subspace spanned by $|0\rangle^{\otimes N}$ 
and $|1\rangle^{\otimes N}$; only even-parity outcomes are retained. 
Tracing out the target system yields a post-selected control state,
\begin{equation}
\rho^{(1)} = \frac{\mathrm{Tr}_T[(\mathbb{I}_C \otimes P_T)\,U(\rho^{\otimes 2})
U^\dagger(\mathbb{I}_C \otimes P_T)]}{p_1},
\end{equation}
where $p_1$ is the corresponding success probability. This step suppresses 
bit-flip errors while preserving GHZ-type coherence; phase errors are 
subsequently addressed by the hashing protocol in Step~3.

\paragraph*{Step 3: Iteration and hashing.}
The above procedure can be iterated $r$ times, producing a sequence of 
post-selected states with success probabilities $\{p_i\}$. After sufficient 
purification, a multipartite hashing protocol is applied~\cite{murao_pra_1998}. 
This yields a lower bound on the distillable GHZ entanglement,
\begin{equation}
\mathcal{E}^{\mathrm{GHZ}}_D \;\ge\;
\max_{r \ge 0}
\left[
R_{\mathrm{hash}}(r)\prod_{i=1}^r p_i
\right],
\end{equation}
where the hashing rate $R_{\mathrm{hash}}$, defined in Step~4 below, 
quantifies the asymptotic fraction of perfect GHZ states extractable 
per copy via one-way classical communication.

\paragraph*{Step 4: GHZ stabilizer representation.}
The hashing rate is given by
\begin{equation}
R_{\mathrm{hash}} = 1 - H(P_\mu) - H(P_\nu),
\end{equation}
where $P_{\mu\nu} = \langle \mu\nu | \rho | \mu\nu \rangle$ are 
populations in the GHZ stabilizer basis
\begin{equation}
|\mu\nu\rangle
= \frac{1}{\sqrt{2}}
\left(
|\mu,0\rangle + (-1)^\nu |\overline{\mu,0}\rangle
\right),
\end{equation}
with $\mu \in \{0,1\}^{N-1}$ labeling phase errors and $\nu \in \{0,1\}$ 
labeling bit-flip errors. The quantities $H(P_\mu)$ and $H(P_\nu)$ are 
the Shannon entropies of the corresponding marginal distributions over 
phase and bit-flip errors, respectively.

Operationally, the bilateral CNOT and postselection steps convert 
multipartite phase correlations into detectable parity information, 
suppressing bit-flip errors. The hashing stage then addresses residual 
phase errors, extracting asymptotically perfect GHZ states at a finite 
rate whenever the total entropy lies below the stabilizer threshold.

\subsection{GHZ distillation protocol for the reduced density matrix in Bethe-basis}

The input state for the recurrence protocol, expressed in the restricted 
Bethe basis $\{ \ket{\Psi_0}, \ket{\Psi_1(q=0)}, \ket{\Psi_{N-1}(q=0)}, 
\ket{\Psi_N} \}$, where $q=0$ selects the zero-momentum sector, has the 
block-diagonal form
\begin{equation}
  \rho^{(0)} =
  \begin{pmatrix}
      \frac{u}{2} & 0 & 0 & v \\
      0 & \frac{w}{2} & 0 & 0 \\
      0 & 0 & \frac{w}{2} & 0 \\
      v^* & 0 & 0 & \frac{u}{2}
  \end{pmatrix},
\end{equation}
where $u, w \ge 0$ and $v \in \mathbb{C}$, subject to the normalization 
condition $u + w + 2\,\mathrm{Re}\,v = 1$.

The distillation step follows a standard multipartite recurrence 
protocol~\cite{dur_ropp_2007,dur_prl_2003}. Two identical copies of the 
state are taken, one as the control and the other as the target. Each party 
applies a local CNOT gate from control to target, followed by a global 
even-parity projective measurement on the target register onto the subspace 
spanned by $\ket{0}^{\otimes N}$ and $\ket{1}^{\otimes N}$. Successful 
outcomes are post-selected, yielding a renormalized control state.

In the Bethe basis, the action of the global CNOT operation simplifies 
considerably. The product states $\ket{0}^{\otimes N}$ and $\ket{1}^{\otimes N}$ 
are mapped deterministically as $\ket{0}^{\otimes N}\ket{0}^{\otimes N} 
\to \ket{0}^{\otimes N}\ket{0}^{\otimes N}$ and $\ket{1}^{\otimes N}
\ket{0}^{\otimes N} \to \ket{1}^{\otimes N}\ket{1}^{\otimes N}$, while 
the one- and $(N-1)$-excitation sectors contribute amplitudes suppressed 
by $1/\sqrt{N}$, since the overlap of a single-magnon Bethe state with 
a product state scales as $1/\sqrt{N}$~\cite{izyu}. 
Repeated post-selection thus progressively filters out Dicke-type 
contributions in large-$N$ many-body systems.

\subsubsection{Recurrence relations}

After one successful iteration, the unnormalized density matrix 
$\sigma_{(l+1)}$ retains the same block structure as $\rho_{(l)}$, 
with coefficients obeying
\begin{align}
  u_{l+1} &= u_l^2, \\
  w_{l+1} &= \frac{w_l^2}{N}, \\
  v_{l+1} &= 2(\mathrm{Re}\,v_l)^2,
\end{align}
where the factor of 2 in $p_{l+1}$ below arises from the two equal 
diagonal blocks of $u/2$ in $\rho^{(0)}$. In the thermodynamic limit 
$N \to \infty$, the recurrence $w_{l+1} = w_l^2/N$ causes the Dicke 
contribution $w_l$ to be exponentially suppressed and can be neglected. 
The remaining coefficients then admit closed-form solutions,
\begin{align}
  u_r &= u^{2^r}, \\
  v_r &= 2^{2^r-1}\,(\mathrm{Re}\,v)^{2^r-1}\,v,
\end{align}
which can be verified by direct substitution into the recurrence relations.
The success probability at the $(l+1)$-th iteration is
\begin{equation}
  p_{l+1} = \mathrm{Tr}\,\sigma_{(l+1)} = 2u_l^2,
\end{equation}
leading to the cumulative success probability
\begin{equation}
  \prod_{i=1}^r p_i = u^{2^{r+1}-2}.
\end{equation}
After normalization, the control state at the $r$-th iteration approaches 
the GHZ-diagonal form
\begin{equation}
  \rho^{(r)} =
  \begin{pmatrix}
      \frac{1}{2} & 0 & 0 & v_r \\
      0 & 0 & 0 & 0 \\
      0 & 0 & 0 & 0 \\
      v_r^* & 0 & 0 & \frac{1}{2}
  \end{pmatrix},
 \label{eq:rho_r}
\end{equation}
confirming that the Dicke-type contributions vanish asymptotically and 
the state converges to the GHZ-diagonal subspace.

\subsubsection{GHZ stabilizer projections and rate}

Since $\rho^{(r)}$ is supported entirely on the $\mu = 0\ldots0$ phase 
sector, only two GHZ stabilizer basis states~\cite{gottesman_pra_1996} contribute non-vanishing weights,
\[
(\mu,\nu) = (0\ldots0,\,0),\quad (0\ldots0,\,1).
\]
Their probabilities are
\begin{align}
  P^{(r)}_{0\ldots0,0} &= \tfrac{1}{2}\left[1 + 
    (2\,\mathrm{Re}\,v)^{2^r}\right], \\
  P^{(r)}_{0\ldots0,1} &= \tfrac{1}{2}\left[1 - 
    (2\,\mathrm{Re}\,v)^{2^r}\right].
\end{align}
Consequently, the marginal over phase errors is trivially concentrated, 
$P^{(r)}_\mu = 1$, contributing zero Shannon entropy $H(P_\mu) = 0$, 
while the bit-flip marginal is 
$P^{(r)}_\nu = \{P^{(r)}_0, P^{(r)}_1\}$.
Following~\cite{dur_2000,dur_ropp_2007}, the achievable 
lower bound on the distillable GHZ entanglement is
\begin{equation}
  \mathcal{E}^{\mathrm{GHZ}}_D =
  \max_{r \ge 0}
  \left[
    u^{\,2^{r+1}-2}
    \left(
      1 - h\left(P^{(r)}_0,\, P^{(r)}_1\right)
    \right)
  \right],
\end{equation}
where $h$ denotes the binary Shannon entropy~\cite{nielsen2000quantum}. 
This expression highlights the competition between the exponential 
purification of GHZ coherence, captured by the growth of 
$(2\,\mathrm{Re}\,v)^{2^r}$ toward unity, and the exponentially 
decreasing cumulative success probability $u^{2^{r+1}-2}$, which 
together determine the optimal distillation depth~$r$.

\section{Geometric measure for multipartite entanglement for the generalized GHZ state and the generalized $W$ state}\label{AppD}

\subsection{GME for the evolved GHZ state}

The geometric measure of genuine multipartiteentanglement (GME) for a pure
$N$-qubit state $\ket{\psi}$ is defined as~\cite{wei_pra_2003}
\begin{equation}
E_G(\ket{\psi}) \equiv 1 - \Lambda(\ket{\psi}),
\end{equation}
where
\begin{equation}
\Lambda(\ket{\psi}) \equiv
\max_{\ket{\phi} \in \mathrm{BS}} \vert \langle \phi|\psi \rangle \vert^2 ,
\end{equation}
and $\mathrm{BS}$ denotes the set of all pure biseparable states over all possible
bipartitions of the $N$ parties.
For pure states, $E_G$ satisfies all axioms of a genuine multipartite entanglement
monotone~\cite{wei_pra_2003,hubener_pra_2009}.

For a mixed state $\rho$, the geometric measure is extended via the convex-roof
construction~\cite{wei_pra_2003,bastin_prl_2009,das_pra_2016}
\begin{equation}
E_G(\rho)
=
\inf_{\{p_i,\ket{\psi_i}\}}
\sum_i p_i\, E_G(\ket{\psi_i}),
\qquad
\rho = \sum_i p_i \ket{\psi_i}\!\bra{\psi_i}.
\end{equation}

We consider a class of $N$-qubit mixed states on a restricted Bethe basis
$\{ \ket{\Psi_0}, \ket{\Psi_1(q=0)}, \ket{\Psi_{N-1}(q=0)}, \ket{\Psi_N} \}$,
with density matrix
\begin{equation}
  \rho =
  \begin{pmatrix}
      \frac{u}{2} & 0 & 0 & v \\
      0 & \frac{1-u}{2} & 0 & 0 \\
      0 & 0 & \frac{1-u}{2} & 0 \\
      v & 0 & 0 & \frac{u}{2}
  \end{pmatrix},
  \qquad 0 \le u \le 1 .
\end{equation}

Due to orthogonality of different magnon-number sectors, $\rho$ admits a convex
decomposition into three mutually orthogonal components:
(i) the zero- and $N$-magnon sector,
(ii) the one-magnon sector,
and (iii) the $(N-1)$-magnon sector.
Explicitly,
\begin{equation}
\rho = u\, \rho_{0N} + \frac{1-u}{2}\, \rho_{1} + \frac{1-u}{2}\, \rho_{N-1},
\end{equation}
where
\begin{align}
\rho_{0N} &=
\begin{pmatrix}
   \frac{1}{2} & \frac{v}{u} \\
   \frac{v^*}{u} & \frac{1}{2}
\end{pmatrix}, \\
\rho_{1} &= \ket{\Psi_1(q=0)}\!\bra{\Psi_1(q=0)}, \\
\rho_{N-1} &= \ket{\Psi_{N-1}(q=0)}\!\bra{\Psi_{N-1}(q=0)} .
\end{align}

Since these sectors are supported on mutually orthogonal subspaces, the convex-roof
optimization can be analyzed sector-wise. In particular, the geometric GME of
$\rho$ is determined by the entanglement properties of the GHZ-like
$\rho_{0N}$ block and the Dicke-type one- and $(N-1)$-magnon states, which are
known to possess genuine multipartite entanglement.


We consider the normalized zero--$N$ magnon block
\begin{equation}
\rho_{0N}
=
\begin{pmatrix}
\frac{1}{2} & \frac{v}{u} \\
\frac{v}{u} & \frac{1}{2}
\end{pmatrix},
\qquad |v|\le \frac{u}{2},
\label{eq:rho_0N}
\end{equation}
written in the basis $\{\ket{\Psi_0},\ket{\Psi_N}\}
=
\{\ket{0}^{\otimes N},\ket{1}^{\otimes N}\}$.
This state is supported entirely in the GHZ subspace. Any pure state in this subspace can be parameterized as
\begin{equation}
\ket{\psi(\theta,\phi)}
=
\cos\theta\,\ket{\Psi_0}
+
e^{i\phi}\sin\theta\,\ket{\Psi_N},
\qquad
0\le\theta\le\frac{\pi}{2}.
\end{equation}
For such states, the maximal overlap with the set of biseparable states is
attained by product states aligned with either
$\ket{\Psi_0}$ or $\ket{\Psi_N}$. Hence
\begin{equation}
\Lambda(\ket{\psi})
=
\max_{\ket{\phi}\in\mathrm{BS}}
 \vert \langle \phi|\psi \rangle \vert^2
=
\max\{\cos^2\theta,\sin^2\theta\}.
\end{equation}
The geometric measure of genuine multipartite entanglement is therefore
\begin{equation}
E_G(\ket{\psi})
=
1-\Lambda(\ket{\psi})
=
\min\{\cos^2\theta,\sin^2\theta\}
=
\frac{1}{2}\left(1-|\cos2\theta|\right).
\label{eq:EG-pure}
\end{equation}

The eigenvalues of the state in Eq.~\eqref{eq:rho_0N} are
\begin{equation}
\lambda_{\pm}
=
\frac{1}{2}
\left(
1\pm\sqrt{1-\frac{4v^2}{u^2}}
\right).
\label{eq:eigenvalues}
\end{equation}

The geometric GME for mixed states is defined via the convex roof
\begin{equation}
E_G(\rho_{0N})
=
\inf_{\{p_i,\ket{\psi_i}\}}
\sum_i p_i E_G(\ket{\psi_i}),
\qquad
\rho_{0N}=\sum_i p_i \ket{\psi_i}\!\bra{\psi_i}.
\end{equation}

Let $\ket{\psi_i}$ be pure states of the form
$\ket{\psi(\theta_i,\phi_i)}$.
Using Eq.~\eqref{eq:EG-pure}, we have
\begin{equation}
\sum_i p_i E_G(\ket{\psi_i})
=
\frac{1}{2}
\sum_i p_i
\left(
1-|\cos2\theta_i|
\right).
\label{eq:avgEG}
\end{equation}

The function $f(x)=|x|$ is convex. Hence, by Jensen inequality,
\begin{equation}
\sum_i p_i |\cos2\theta_i|
\ge
\left|
\sum_i p_i \cos2\theta_i
\right|.
\label{eq:jensen}
\end{equation}
Substituting Eq.~\eqref{eq:jensen} into Eq.~\eqref{eq:avgEG}, we obtain
\begin{equation}
\sum_i p_i E_G(\ket{\psi_i})
\ge
\frac{1}{2}
\left(
1-
\left|
\sum_i p_i \cos2\theta_i
\right|
\right).
\label{eq:lowerbound}
\end{equation}

The $x$--component of the Bloch vector of $\rho_{0N}$ is
\begin{equation}
\frac{2|v|}{u}
=
\sum_i p_i \sin2\theta_i \cos\phi_i.
\end{equation}
Since $|\cos\phi_i|\le 1$, the Cauchy--Schwarz inequality implies
\begin{equation}
\left|
\sum_i p_i \sin2\theta_i \cos\phi_i
\right|
\le
\sum_i p_i |\sin2\theta_i|.
\end{equation}
Using $\sin^2 2\theta_i + \cos^2 2\theta_i =1$, the constraint yields
\begin{equation}
\left|
\sum_i p_i \cos2\theta_i
\right|
=
\sqrt{1-\frac{4v^2}{u^2}}.
\label{eq:optimality}
\end{equation}
Equality is achieved by a symmetric decomposition into two pure states
with phases $\phi=0,\pi$.

Substituting Eq.~\eqref{eq:optimality} into Eq.~\eqref{eq:lowerbound}, we obtain
\begin{equation}
E_G(\rho_{0N})
=
\frac{1}{2}
\left(
1-\sqrt{1-\frac{4v^2}{u^2}}
\right).
\end{equation}

The result is independent of system size $N$ because $\rho_{0N}$ is supported
entirely within the GHZ subspace. The convex-roof minimum is computed by a
two-element ensemble, consistent with general results for rank--$2$ states.

\subsection{Geometric entanglement of $W$ states}

\paragraph*{Geometric collective multipartite entanglement:} For permutation-invariant states such as the generalized $W$ state, the maximal overlap is
achieved by symmetric product states \cite{wei_jmp_2010}. Thus it suffices to consider product states of the form
\begin{equation}
\ket{\phi(\theta)}
=
\left(
\cos\theta\,\ket{0}
+
\sin\theta\,\ket{1}
\right)^{\otimes N},
\qquad
0\le\theta\le\frac{\pi}{2}.
\end{equation}

The overlap between $\ket{\Psi_1(q=0)}$ and $\ket{\phi(\theta)}$ is
\begin{equation}
\braket{\phi(\theta)|\Psi_1(q=0)}
=
\frac{1}{\sqrt{N}}
\sum_{j=1}^N
\sin\theta\,(\cos\theta)^{N-1}
=
\sqrt{N}\,
\sin\theta\,(\cos\theta)^{N-1}.
\end{equation}
Hence
\begin{equation}
|\braket{\phi(\theta)|\Psi_1(q=0)}|^2
=
N\,\sin^2\theta\,(\cos^2\theta)^{N-1}.
\label{eq:overlap}
\end{equation}

To maximize this expression, we differentiate with respect to $\theta$:
\begin{equation}
\frac{d}{d\theta}
\left[
\ln
\left(
\sin^2\theta\,(\cos^2\theta)^{N-1}
\right)
\right]
=
2\cot\theta
-
2(N-1)\tan\theta
=0.
\end{equation}
This yields the optimal angle
\begin{equation}
\sin^2\theta = \frac{1}{N},
\qquad
\cos^2\theta = \frac{N-1}{N}.
\label{eq:optimaltheta}
\end{equation}

Substituting Eq.~\eqref{eq:optimaltheta} into Eq.~\eqref{eq:overlap}, we obtain
\begin{equation}
\Lambda(\ket{\Psi_1(q=0)})
=
\left(
\frac{N-1}{N}
\right)^{N-1}.
\label{eq:lambdaWN}
\end{equation}

Therefore, the geometric measure of genuine multipartite entanglement is
\begin{equation}
E_G(\ket{\Psi_1(q=0)})
=
1-
\left(
\frac{N-1}{N}
\right)^{N-1}.
\label{eq:EGWN}
\end{equation}

In the thermodynamic limit,
\begin{equation}
\lim_{N\to\infty}
\left(
\frac{N-1}{N}
\right)^{N-1}
=
e^{-1},
\end{equation}
we obtain
\begin{equation}
\lim_{N\to\infty}
E_G(\ket{\Psi_1(q=0)})
=
1-\frac{1}{e} \approx 0.6321.
\end{equation}

The maximization over fully separable states suffices because of
permutation symmetry and the convexity of the biseparable set.
 Unlike GHZ states, whose geometric GME is $1/2$ independent of $N$,
the $W$ state exhibits a strictly larger and size-dependent GME.
The finite asymptotic value reflects the robustness of $W$-type
entanglement under particle loss.
\vspace{5mm}

\paragraph*{Geometric genuine mutipartite entanglement:} We consider a bipartition $A|B$ with $|A|=m$ and $|B|=N-m$.
Separating the sum according to whether the excitation lies in $A$ or in $B$, the state can be written as
\begin{equation}
\ket{\Psi_1(q=0)}\equiv |W_N\rangle 
=
\frac{1}{\sqrt{N}}
\left(
\sum_{i\in A} |i\rangle_A |0\rangle_B
+
\sum_{j\in B} |0\rangle_A |j\rangle_B
\right).
\end{equation}

Define the normalized states
\begin{equation}
|W_m\rangle_A
=
\frac{1}{\sqrt{m}}
\sum_{i\in A} |i\rangle_A,
\qquad
|W_{N-m}\rangle_B
=
\frac{1}{\sqrt{N-m}}
\sum_{j\in B} |j\rangle_B.
\end{equation}

Then the state becomes
\begin{equation}
|W_N\rangle
=
\sqrt{\frac{m}{N}}
\, |W_m\rangle_A |0\rangle_B
+
\sqrt{\frac{N-m}{N}}
\, |0\rangle_A |W_{N-m}\rangle_B.
\end{equation}

The two terms above are orthogonal because they belong to different excitation-number sectors of subsystem $A$:
the first term has one excitation in $A$, whereas the second has zero excitations in $A$.
Therefore the decomposition is already in Schmidt form.
The corresponding Schmidt coefficients are
\begin{equation}
\lambda_1^2 = \frac{m}{N},
\qquad
\lambda_2^2 = \frac{N-m}{N}.
\end{equation}

The largest Schmidt coefficient for a fixed bipartition is
\begin{equation}
\lambda_{\max}^2(A|B)
=
\max\!\left(
\frac{m}{N},
\frac{N-m}{N}
\right).
\end{equation}

We now maximize over all possible bipartitions, i.e. over $m=1,2,\dots,N-1$.
Since $\max(m,N-m)$ is largest when one subsystem contains a single site,
the global maximum is achieved at $m=1$ (or equivalently $m=N-1$), yielding
\begin{equation}
\lambda_{\max}^2
=
\frac{N-1}{N}.
\end{equation}

Using the definition of the geometric measure of GME
we obtain
\begin{equation}
E_G(\ket{\Psi_1(q=0)})
=
1-\frac{N-1}{N}
=
\frac{1}{N}.
\end{equation}

\subsection{ Geometric entanglement of the two-magnon sector}

\paragraph*{Geometric collective multipartite entanglement:} We consider the normalized two-magnon state on a periodic chain,
\begin{equation}
\ket{\Psi_2 (q=0)}
=
\sum_{x_1,x_2}
\frac{\sqrt{3}}{N^2}\,(x_2-x_1)\,
\ket{x_1,x_2},
\label{eq:psi2}
\end{equation}
which is translationally invariant  in the
thermodynamic limit. The state $\ket{\Psi_2 (q=0)}$ defined in Eq.~\eqref{eq:psi2} is not fully permutation
invariant due to the position-dependent weight $(x_2-x_1)$. However, the restriction follows from a variational argument as follows. We consider a general ansatz product state~\cite{wei_jmp_2010}
\begin{equation}
|\phi\rangle
=
\bigotimes_{k=1}^N
(\alpha_k |0\rangle+\beta_k |1\rangle),
\qquad
|\alpha_k|^2+|\beta_k|^2=1.
\end{equation}
The overlap reads
\begin{equation}
\langle\phi|\Psi_2 (q=0)\rangle
=
\frac{\sqrt{3}}{N^2}
\sum_{x_1,x_2}
(x_2-x_1)\,
\beta_{x_1}\beta_{x_2}
\prod_{k\neq x_1,x_2}\alpha_k = \frac{\sqrt{3}}{N^2} P
\sum_{x_1,x_2}
(x_2-x_1)
\frac{\beta_{x_1}}{\alpha_{x_2}}
\frac{\beta_{x_1}}{\alpha_{x_2}},
\label{eq:overlap_general}
\end{equation}
where $P = \prod_{k=1}^N \alpha_k$. Introducing the variables $
z_k=\frac{\beta_k}{\alpha_k}$,
$\alpha_k={(1+z_k^2)}^{-1/2}$
we get
\begin{equation}
P=\prod_{k=1}^N \frac{1}{\sqrt{1+z_k^2}},
\end{equation}
and the overlap becomes
\begin{equation}
F(\{z_k\})
= \frac{\sqrt{3}}{N^2}
\left(
\prod_{k=1}^N \frac{1}{\sqrt{1+z_k^2}}
\right)
\sum_{x_1,x_2}(x_2-x_1)z_{x_1} z_{x_2} .
\label{eq:F_functional}
\end{equation}

By defining the symmetric matrix
\begin{equation}
K_{ij}
=
\begin{cases}
|i-j|, & i\neq j,\\
0, & i=j .
\end{cases}
\end{equation}

we obtain the following functional
\begin{equation}
F_N(z_1,\dots,z_N)
=
\frac{\displaystyle \sum_{i,j}^N K_{ij} z_{i} z_{j}}
{\displaystyle \prod_{k=1}^N \sqrt{1+z_k^2}},
\label{eq:122}
\end{equation}
where, $K$ is real, symmetric, and circulant (defined over a periodic boundary condition), $K_{ij} = f(i-j)$ with periodic boundary conditions, $K_{ij} \ge 0$. Our goal is to determine the configuration $(z_1,\dots,z_N)$ that maximizes $F_N$ in the thermodynamic limit.

Since $K$ is circulant, it is diagonalized by discrete Fourier modes:
\begin{equation}
z_j = \sum_{q} c_q e^{i q j},
\end{equation}
where $q = 2\pi n/N$. Then
\begin{equation}
z^T K z
=
N \sum_q \lambda(q) |c_q|^2,
\end{equation}
where $\lambda(q)$ is the Fourier transform of $f$. Because $K_{ij} \ge 0$,
its Fourier transform satisfies~\cite{davis}
\begin{equation}
\lambda(0) > \lambda(q)
\quad \text{for } q \neq 0.
\end{equation}
Therefore the numerator in Eq.~\eqref{eq:122} is maximized when, $z_j = \text{constant}$. 

Now we consider the denominator  in Eq.~\eqref{eq:122}:
\begin{equation}
\frac{1}{N} \sum_{k=1}^N \log(1+z_k^2).
\end{equation}
As $N \to \infty$, this becomes the  average
of $\log(1+z^2)$ over the distribution of site values.
For any configuration we define
\begin{equation}
m_2 = \frac{1}{N} \sum_{k=1}^N z_k^2.
\end{equation}
Since $\log(1+t)$ is strictly convex for $t \ge 0$, Jensen's inequality yields $
\frac{1}{N} \sum_{k=1}^N \log(1+z_k^2)
\ge
\log(1+m_2)$,
with equality if and only if all $z_k^2$ are equal. Hence, for fixed second moment $m_2$,
the denominator per site is minimized by a uniform configuration.

\vspace{1em}
Therefore  we  restrict the optimization to symmetric product states of the form
\begin{equation}
\ket{\phi(\theta)}
=
(\cos\theta\,\ket{0}+\sin\theta\ket{1})^{\otimes N},
\qquad
0\le\theta\le\frac{\pi}{2}.
\end{equation}

The overlap is
\begin{equation}
\braket{\phi(\theta)|\Psi_2(q=0)}
=
\frac{\sqrt{3}}{N^2}
\sum_{x_1<x_2}
(x_2-x_1)
\sin^2\theta
(\cos\theta)^{N-2}.
\end{equation}

Using translational invariance, the sum depends only on separations
$r=(x_2-x_1)$,
\begin{equation}
\sum_{x_1,x_2} (x_2-x_1)
=
N \sum^{N-1}_{r=1} r = \frac{N^2(N-1)}{2}.
\end{equation}
Hence, for large $N$,
\begin{equation}
|\braket{\phi(\theta)|\Psi_2(q=0)}|^2
=
\frac{3}{4}
{N^2}
\sin^4\theta
(\cos^2\theta)^{N-2}
\label{eq:overlap2}
\end{equation}

Maximizing Eq.~\eqref{eq:overlap2} with respect to $\theta$ is equivalent to
maximizing
\[
f(\theta)
=
\sin^4\theta
(\cos^2\theta)^{N-2}.
\]
Setting $\partial_\theta \ln f(\theta)=0$ yields
\begin{equation}
4\cot\theta - 2(N-2)\tan\theta = 0,
\end{equation}
which gives the optimal angle
\begin{equation}
\sin^2\theta = \frac{2}{N},
\qquad
\cos^2\theta = 1-\frac{2}{N}.
\label{eq:opt2}
\end{equation}

Substituting Eq.~\eqref{eq:opt2} into Eq.~\eqref{eq:overlap2},
\begin{equation}
\Lambda(\ket{\Psi_2(q=0)})
=
\frac{3}{4} N^2
\left(1-\frac{2}{N}\right)^{N-2} \times \frac{4}{N^2}
=
3
\left(1-\frac{2}{N}\right)^{N-2}.
\end{equation}

Taking the thermodynamic limit,
\begin{equation}
\lim_{N\to\infty}
\left(1-\frac{2}{N}\right)^{N-2}
=
e^{-2},
\end{equation}
we obtain
\begin{equation}
\Lambda(\ket{\Psi_2(q=0)})
=
\frac{3}{e^2}.
\end{equation}

Therefore,
\begin{equation}\lim_{N\to\infty}
E_G(\ket{\Psi_2(q=0)})
=
1-\frac{3}{e^2} \approx 0.5940.
\end{equation}
 The scaling $\sin^2\theta\sim 2/N$ reflects the two-magnon nature of the
state.
Compared to the generalized $W$ state, the overlap decreases faster with $N$,
leading to a larger geometric GME.
 This result is consistent with the general behavior of symmetric
Dicke states with more than one excitations \cite{hubener_pra_2009}.

\paragraph*{Geometric genuine multipartite entanglement:}

We consider a bipartition $A|B$ with $|A|=m$ and $|B|=N-m$. Because the state contains exactly two excitations, the reduced density
matrix $\rho_A$ decomposes into orthogonal excitation sectors
\begin{equation}
\rho_A
=
\rho_{AA}
\oplus
\rho_{AB}
\oplus
\rho_{BB},
\end{equation}
corresponding respectively to  
(a) two excitations in $A$,  
(b) one excitation in $A$,  
(c) zero excitations in $A$.

This block decomposition follows from particle-number conservation: sectors with different excitation numbers in subsystem $A$ are orthogonal and therefore do not mix in $\rho_A$.

The traces of the outer blocks define the probabilities
\begin{equation}
p_{AA}=\mathrm{tr}\,\rho_{AA},
~~~~
p_{BB}=\mathrm{tr}\,\rho_{BB},
\end{equation}
which represent the probabilities that both magnons reside in $A$ or both reside in $B$, respectively. The one–excitation block $\rho_{AB}$ has at most two nonzero
eigenvalues, which we denote by$\lambda_1$ and $\lambda_2$.

Since the blocks act on orthogonal sectors, the spectrum of $\rho_A$ is the union $ 
\{p_{AA}\}
\cup
\{\lambda_1,\lambda_2\}
\cup
\{p_{BB}\}$. Therefore the largest Schmidt coefficient across any cut is
\begin{equation}
\lambda_{\max}
=
\max\big(
p_{AA},
p_{BB},
\lambda_1,
\lambda_2
\big),
\label{eq: spectrum}
\end{equation}
with the normalization condition
\begin{equation}
p_{AA} + p_{BB} + \lambda_1 + \lambda_2 = 1,
\end{equation}
which follows from $\mathrm{tr}~\rho_A=1$.


We now specialize to the cut where $A$ consists of a single site. A single site cannot host two excitations simultaneously, hence
\begin{equation}
p_{AA}=0.
\end{equation}

The quantity $p_{AB}$ corresponds to configurations in which exactly one magnon occupies site $x_1\in A$ while the second magnon resides at some $x_2\neq x_1$ in $B$. Therefore,
\begin{eqnarray}
p_{AB} =
\sum_{x_2\neq x_1}
\left|
\frac{\sqrt{3}}{N^2}
(x_2-1)
\right|^2 = \frac{1}{4N} + \frac{1}{2N^3}.
\label{eq:p_ab}
\end{eqnarray}

Now, subsystem $A$ has dimension one in the single–excitation sector, $\rho_{AB}$ is rank one. Therefore,
\begin{equation}
\lambda_1 = p_{AB},
\qquad
\lambda_2 = 0.
\end{equation}

The remaining probability weight lies in the zero–excitation sector
\begin{equation}
p_{BB}
=
1 - p_{AB}.
\end{equation}

Thus  from Eq.~\eqref{eq: spectrum} the full spectrum of $\rho_A$ now becomes
\begin{equation}
\{\, 0, p_{AB}, 1-p_{AB} \}.
\end{equation}

Hence
\begin{equation}
\lambda_{\max}
=
\max\big(p_{AB},1-p_{AB}\big).
\end{equation}

From Eq.~\eqref{eq:p_ab}, in the large $N$ limit, the dominant eigenvalue is
\begin{equation}
\lambda_{\max}
=
1
-
\frac{1}{4N}
-
\frac{1}{2N^3}.
\end{equation}

Thus in the thermodynamic limit $\lambda_{\max}\to 1$, implying that the microscopic $(1|N-1)$ cut dominates the maximization over all bipartitions. Using the definition of the geometric measure of GME,
$E_G = 1-\lambda_{\max}^2$,
we expand perturbatively in $1/N$ and obtain
\begin{equation}
E_G
=
\frac{1}{2N}
-
\frac{1}{16 N^2}
+
\mathcal{O}(N^{-3}).
\end{equation}

Therefore the geometric GME vanishes in the thermodynamic limit, with leading finite-size correction proportional to $(2N)^{-1}$.

\subsection{Exactness of the convex roof for the generalized 
$W$-state dynamics}

Let $\rho = \sum_{k=0}^{3} w_k \ket{\Psi_k}\bra{\Psi_k}$ be 
the time-evolved state of the generalized $W$ state, diagonal 
in the Bethe basis $\{\ket{\Psi_0}, \ket{\Psi_1(q=0)}, 
\ket{\Psi_2(q=0^+)}, \ket{\Psi_2(q=\pi^-)}\}$, with $w_k > 0$ 
and $\sum_k w_k = 1$. In the thermodynamic limit $N \to \infty$, 
the convex roof of the geometric collective multipartite 
entanglement (CME) is achieved exactly by the spectral 
decomposition
\begin{eqnarray}
E_G^{\mathrm{roof}}(\rho) = \sum_{k=0}^{3} w_k\, 
E_G(\ket{\Psi_k}),
\end{eqnarray}
where $E_G(\ket{\Psi_k}) = 1 - \Lambda^2_{\max}(\ket{\Psi_k})$ 
and $\Lambda_{\max}(\ket{\Psi_k}) = 
\max_{\ket{s}\in\mathrm{sep}}|\braket{s|\Psi_k}|^2$ is the 
maximum overlap with any fully separable state.

\begin{proof}
The convex roof is defined as
\begin{eqnarray}
E_G^{\mathrm{roof}}(\rho) = 
\inf_{\{p_j, \ket{\phi_j}\}} 
\sum_j p_j\, E_G(\ket{\phi_j}),
\end{eqnarray}
where the infimum is over all pure state decompositions 
$\rho = \sum_j p_j \ket{\phi_j}\bra{\phi_j}$. We show 
that in the thermodynamic limit every pure state 
decomposition yields the same average geometric CME, 
so the infimum is achieved everywhere and equals the 
spectral decomposition value.

\medskip
\noindent\textbf{(a)} Any pure state decomposition of $\rho$ is related to the 
spectral decomposition by a unitary matrix 
$U$~\cite{nielsen2000quantum}
\begin{eqnarray}
\sqrt{p_j}\,\ket{\phi_j} = \sum_{k=0}^{3} U_{jk} 
\sqrt{w_k}\,\ket{\Psi_k},
\end{eqnarray}
so each $\ket{\phi_j}$ is a superposition:
\begin{eqnarray}
\ket{\phi_j} = \alpha_j \ket{\Psi_0} 
+ \beta_j \ket{\Psi_1(q=0)} 
+ \gamma_j \ket{\Psi_2(q=0^+)} 
+ \delta_j \ket{\Psi_2(q=\pi^-)},
\label{eq:phij}
\end{eqnarray}
with $|\alpha_j|^2 + |\beta_j|^2 + |\gamma_j|^2 + 
|\delta_j|^2 = 1$. Unitarity of $U$ implies the sum 
rules, valid for any decomposition
\begin{align}
\sum_j p_j |\alpha_j|^2 = w_0, \quad
\sum_j p_j |\beta_j|^2  = w_1, \quad
\sum_j p_j |\gamma_j|^2 = w_2(0^+), \quad
\sum_j p_j |\delta_j|^2 = w_2(\pi^-).
\label{eq:sumrules}
\end{align}

\medskip
\noindent\textbf{(b)} The four Bethe states belong to distinct magnetization 
sectors: $\ket{\Psi_0}$ to $l=0$, $\ket{\Psi_1(q=0)}$ 
to $l=1$, and $\ket{\Psi_2(q=0^+)}$, 
$\ket{\Psi_2(q=\pi^-)}$ to $l=2$. For any separable 
state $\ket{s} = (\cos\theta\ket{0}+\sin\theta\ket{1})^{
\otimes N}$, the overlaps with states in different sectors 
are
\begin{align}
A^{(s)} &\equiv \braket{s|\Psi_0} 
= \cos^N\theta,\\
B^{(s)} &\equiv \braket{s|\Psi_1(q=0)} 
= \sqrt{N}\sin\theta\cos^{N-1}\theta,\\
C^{(s)}_{0^+} &\equiv \braket{s|\Psi_2(q=0^+)} 
= \frac{\sqrt{3}(N-1)}{2}
\sin^2\theta\cos^{N-2}\theta.
\end{align}
The cross terms between different sectors scale as
\begin{align}
\lim_{N \rightarrow \infty}A^{(s)*}B^{(s)} &= 
\sqrt{N}\sin\theta\cos^{2N-1}\theta 
= 0,\\
\lim_{N \rightarrow \infty}A^{(s)*}C^{(s)}_{0^+} &= 
\frac{\sqrt{3}(N-1)}{2}
\sin^2\theta\cos^{2N-2}\theta 
= 0,\\
\lim_{N \rightarrow \infty}B^{(s)*}C^{(s)}_{0^+} &= 
\frac{\sqrt{3N}(N-1)}{2}
\sin^3\theta\cos^{2N-3}\theta 
= 0,
\end{align}
uniformly for all $\theta\in[0,\pi/2]$, since any 
polynomial in $N$ multiplied by $\cos^{cN}\theta$ for 
$c > 0$ vanishes as $N\to\infty$ for all 
$\theta\in(0,\pi/2]$, and at $\theta=0$ all cross terms 
are trivially zero. The same argument applies to cross 
terms involving $C^{(s)}_{\pi^-}$. Therefore, in the thermodynamic limit, the squared overlap 
$|\braket{s|\phi_j}|^2$ factorizes into independent 
magnetization sector contributions
\begin{eqnarray}
\lim_{N \rightarrow \infty} |\braket{s|\phi_j}|^2 \rightarrow
|\alpha_j|^2|A^{(s)}|^2 + |\beta_j|^2|B^{(s)}|^2 
+ |\gamma_j C^{(s)}_{0^+} 
+ \delta_j C^{(s)}_{\pi^-}|^2.
\label{eq:factorized}
\end{eqnarray}

\medskip
\noindent\textbf{(c)} We now show that 
$\Lambda^2_{\max}(\ket{\Psi_2(q=0^+)}) = 
\Lambda^2_{\max}(\ket{\Psi_2(q=\pi^-)})$, 
without needing their explicit values. The two states are
\begin{align}
\ket{\Psi_2(q=0^+)} &= \frac{\sqrt{3}}{N^2}
\sum_{x_1<x_2}(x_2-x_1)\ket{x_1,x_2},\\
\ket{\Psi_2(q=\pi^-)} &= \frac{\sqrt{3}}{N^2}
\sum_{x_1<x_2}(x_2-x_1)(-1)^{x_2-x_1}\ket{x_1,x_2}.
\end{align}
For $\ket{\Psi_2(q=0^+)}$, the maximum overlap with any 
separable state is achieved by the uniform ansatz 
$\ket{s^*_{0^+}} = (\cos\theta^*\ket{0} + 
\sin\theta^*\ket{1})^{\otimes N}$ (see S.I.~IVD). 
For $\ket{\Psi_2(q=\pi^-)}$, we choose the staggered 
ansatz $\ket{s^*_{\pi^-}} = (\cos\theta^*\ket{0} + 
(-1)^k\sin\theta^*\ket{1})^{\otimes N}$ with site-dependent 
phases $\phi_k = k\pi$. Therefore
\begin{eqnarray}
(-1)^{x_2-x_1}\cdot e^{i(\phi_{x_1}+\phi_{x_2})} 
= (-1)^{x_2-x_1}\cdot(-1)^{x_1+x_2} 
= (-1)^{2x_2} = 1,
\end{eqnarray}
so the phase factor cancels completely and the overlap 
reduces to:
\begin{eqnarray}
\braket{s^*_{\pi^-}|\Psi_2(q=\pi^-)} = 
\braket{s^*_{0^+}|\Psi_2(q=0^+)},
\end{eqnarray}
since the staggered ansatz $\ket{s^*_{\pi^-}}$ is related 
to the uniform ansatz $\ket{s^*_{0^+}}$ by a local unitary 
$\sigma^z$ on every other site, which preserves 
separability and the optimal angle $\theta^*$. Therefore
\begin{eqnarray}
\Lambda^2_{\max}(\ket{\Psi_2(q=0^+)}) = 
\Lambda^2_{\max}(\ket{\Psi_2(q=\pi^-)}) 
\equiv \Lambda^2_2.
\label{eq:equal_lambda}
\end{eqnarray}
As $\ket{s^*_{0^+}}$, the overlap with 
$\ket{\Psi_2(q=\pi^-)}$ is suppressed:
\begin{eqnarray}
\lim_{N \rightarrow \infty} |\braket{s^*_{0^+}|\Psi_2(q=\pi^-)}|^2 = 
\mathcal{O}\left(\frac{1}{N^2}\right) 
\rightarrow 0,
\end{eqnarray}
since the alternating phase $(-1)^{x_2-x_1}$ causes 
near-complete cancellation in the sum at the uniform 
ansatz. Similarly $|\braket{s^*_{\pi^-}|\Psi_2(q=0^+)}|^2 
\to 0$. Therefore, maximizing over all separable states
\begin{eqnarray}
\lim_{N \rightarrow \infty} \max_{\ket{s}\in\mathrm{sep}}
|\gamma_j\braket{s|\Psi_2(q=0^+)} + 
\delta_j\braket{s|\Psi_2(q=\pi^-)}|^2
=
(|\gamma_j|^2 + |\delta_j|^2)\Lambda^2_2.
\label{eq:two_magnon_final}
\end{eqnarray}

\medskip
\noindent\textbf{(d)} Combining Steps (b) and (c), the maximum overlap for any 
$\ket{\phi_j}$  in the thermodynamic 
limit becomes
\begin{eqnarray}
\lim_{N \rightarrow \infty} \Lambda^2_{\max}(\ket{\phi_j}) 
= 
|\alpha_j|^2 \Lambda^2_{\max}(\ket{\Psi_0}) 
+ |\beta_j|^2 \Lambda^2_{\max}(\ket{\Psi_1}) 
+ (|\gamma_j|^2 + |\delta_j|^2)\Lambda^2_2,
\label{eq:lambda_factorized}
\end{eqnarray}
where the sectors contribute independently because their 
optimal separable states are distinct and the cross-sector 
overlaps vanish.

\noindent\textbf{(e)} For any pure state decomposition $\{p_j, \ket{\phi_j}\}$ we have
\begin{eqnarray}
\lim_{N \rightarrow \infty} \sum_j p_j E_G(\ket{\phi_j}) 
&=& 1 - \sum_j p_j \Lambda^2_{\max}(\ket{\phi_j})
=1 - \sum_j p_j \Big[
|\alpha_j|^2 \Lambda^2_{\max}(\ket{\Psi_0}) 
+ |\beta_j|^2 \Lambda^2_{\max}(\ket{\Psi_1}) 
+ (|\gamma_j|^2 + |\delta_j|^2)\Lambda^2_2
\Big].
\nonumber\\
\end{eqnarray}
Applying the sum rules from Step~(a), we have
\begin{eqnarray}
\lim_{N \rightarrow \infty} \sum_j p_j E_G(\ket{\phi_j}) 
=
1 - \Big[
w_0 \Lambda^2_{\max}(\ket{\Psi_0}) 
+ w_1 \Lambda^2_{\max}(\ket{\Psi_1}) 
+ (w_2(0^+) + w_2(\pi^-))\Lambda^2_2
\Big]
= \sum_k w_k E_G(\ket{\Psi_k}),
\end{eqnarray}
where the last equality uses 
$E_G(\ket{\Psi_2(q=0^+)}) = E_G(\ket{\Psi_2(q=\pi^-)}) 
= 1 - \Lambda^2_2$. Since this holds for every
pure state decomposition of $\rho$, the infimum in the 
convex roof is achieved by every decomposition equally
\begin{eqnarray}
E_G^{\mathrm{roof}}(\rho) = 
\inf_{\{p_j,\ket{\phi_j}\}} 
\sum_j p_j E_G(\ket{\phi_j}) 
= \sum_k w_k E_G(\ket{\Psi_k}).
\quad\blacksquare
\end{eqnarray}
\end{proof}